\documentclass{article}

\usepackage{microtype}
\usepackage{graphicx}
\usepackage{subcaption}
\usepackage{booktabs}

\usepackage{hyperref}

\usepackage[preprint]{icml2026}

\usepackage{amsmath}
\usepackage{amssymb}
\usepackage{mathtools}
\usepackage{amsthm}
\usepackage{stfloats}

\usepackage[capitalize,noabbrev]{cleveref}

\usepackage[utf8]{inputenc}
\usepackage[T1]{fontenc}
\usepackage{graphicx}
\usepackage{listings}
\usepackage{subcaption}
\usepackage{ulem}
\usepackage{amsmath}
\usepackage{mathtools}
\usepackage{hyperref}
\usepackage{wrapfig}
\usepackage{algorithm}
\setcitestyle{round}
\hypersetup{
   colorlinks=true,
   linkcolor=[RGB]{30, 30, 180},
   citecolor=[RGB]{30, 30, 180},
   urlcolor=magenta,
   pdfborder=0 0 0,
   pdftitle={},
   pdfsubject={}, 
   pdfkeywords={},
   pdfauthor={},%
   pdfstartview=FitH
}
\usepackage[capitalize,noabbrev]{cleveref}
\usepackage[nolist,nohyperlinks]{acronym}
\usepackage{pifont}
\usepackage{booktabs}
\usepackage[table]{xcolor}
\usepackage{svg}
\usepackage[inline]{enumitem}
\usepackage{booktabs}
\usepackage{adjustbox}
\usepackage{multirow}

\usepackage{wrapfig}
\usepackage{array}
\usepackage{colortbl}

\definecolor{codegreen}{rgb}{0,0.6,0}
\definecolor{codegray}{rgb}{0.5,0.5,0.5}
\definecolor{codepurple}{rgb}{0.58,0,0.82}
\definecolor{codeblue}{rgb}{0,0,0.8}

\definecolor{pastelgreen}{HTML}{DFF0D8}
\definecolor{pastelred}{HTML}{F2DEDE}
\definecolor{pastelorange}{HTML}{FCF8E3}

\newcommand{\textbfp}[1]{\vspace{0.5em}\noindent\textbf{#1.}}

\usepackage{tikz}
\newcommand*\circnum[1]{%
  \tikz[baseline=(char.base)]{
    \node[shape=circle,fill=black,inner sep=1pt] (char)
    {\color{white}\footnotesize\textbf{#1}};}}

\usepackage{array}
\usepackage{amsmath}
\usepackage{amssymb}
\usepackage{mathtools}
\usepackage{amsthm}
\usepackage{bm}

\newcommand\Bb{\bm{b}}

\newcommand\Br{\bm{r}}

\newcommand\Bv{\bm{v}}

\newcommand\BA{\bm{A}}
\newcommand\BB{\bm{B}}

\newcommand\BE{\bm{E}}

\newcommand\BR{\bm{R}}

\makeatletter
\newcommand{\dlmf}[1]{%
\citep[%
  \def\nextitem{\def\nextitem{, }}%
  \@for \el:=#1\do{\nextitem\href{http://dlmf.nist.gov/\el}{(\el)}}%
]{olver_nist_2010}%
}
\makeatother

\usepackage[textsize=tiny]{todonotes}

\begin{acronym}[LONGEST]
  \acro{PINC}{Physics-Informed Neural Compression}
\end{acronym}

\icmltitlerunning{
\includegraphics[width=0.75em]{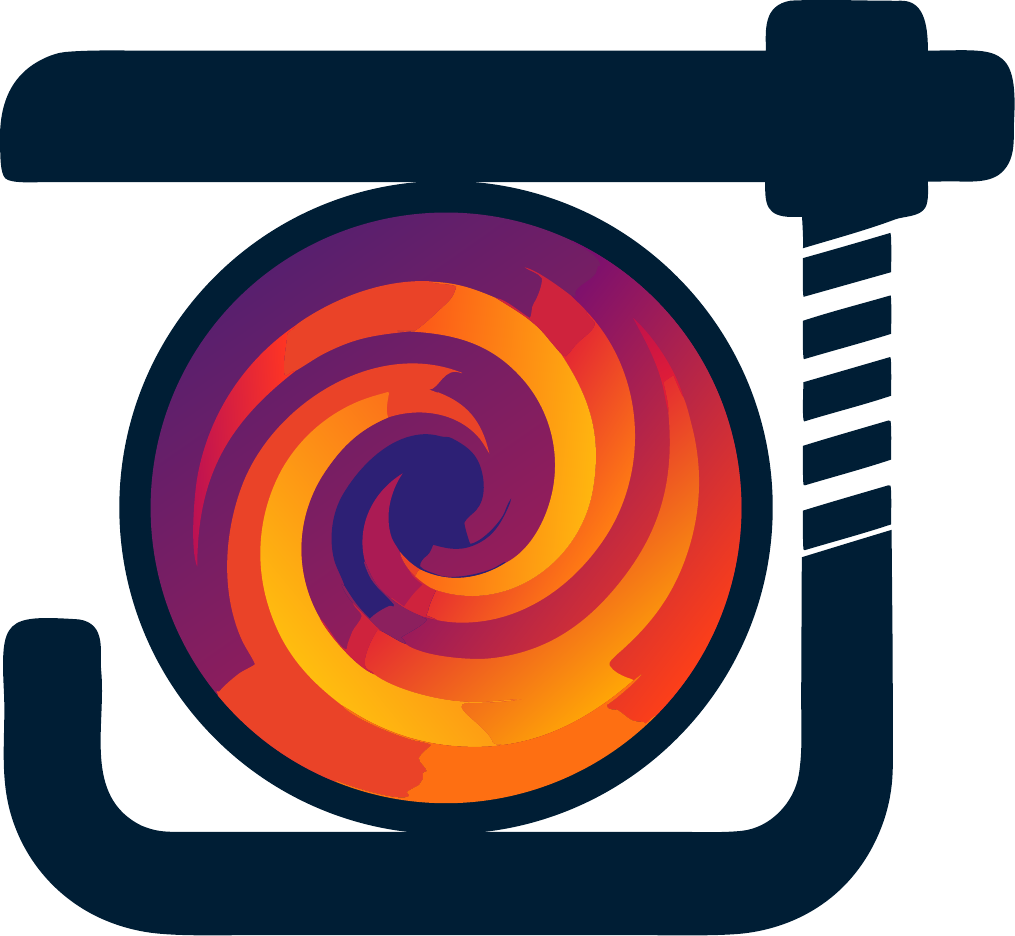}\hspace{0.2em}%
Physics-Informed Neural Compression
}

\begin{document}

\twocolumn[
\icmltitle{
\raisebox{-1.75pt}{\includegraphics[width=1.1em]{figs/pinc_logo.pdf}}\hspace{0.2em}%
Physics-Informed Neural Compression of High-Dimensional Plasma Data
}
  \icmlsetsymbol{equal}{*}
  \begin{icmlauthorlist}
    \icmlauthor{Gianluca Galletti}{jku}
    \icmlauthor{Gerald Gutenbrunner}{jku}
    \icmlauthor{Sandeep S. Cranganore}{jku}
    \icmlauthor{William Hornsby}{ukaea}
    \icmlauthor{Lorenzo Zanisi}{ukaea}
    \icmlauthor{Naomi Carey}{ukaea}
    \icmlauthor{Stanislas Pamela}{ukaea}
    \icmlauthor{Johannes Brandstetter}{jku,emmi}
    \icmlauthor{Fabian Paischer}{jku,emmi}
  \end{icmlauthorlist}

  \icmlaffiliation{jku}{Institute for Machine Learning, JKU Linz}
  \icmlaffiliation{ukaea}{United Kingdom Atomic Energy Authority, Culham campus}
  \icmlaffiliation{emmi}{EMMI AI GmbH, Linz}

  \icmlcorrespondingauthor{Gianluca Galletti}{g.galletti@ml.jku.at}

  \begin{center}
    \href{https://github.com/ml-jku/neural-gyrokinetics}{%
        \raisebox{-0.3em}{\includegraphics[width=1.15em]{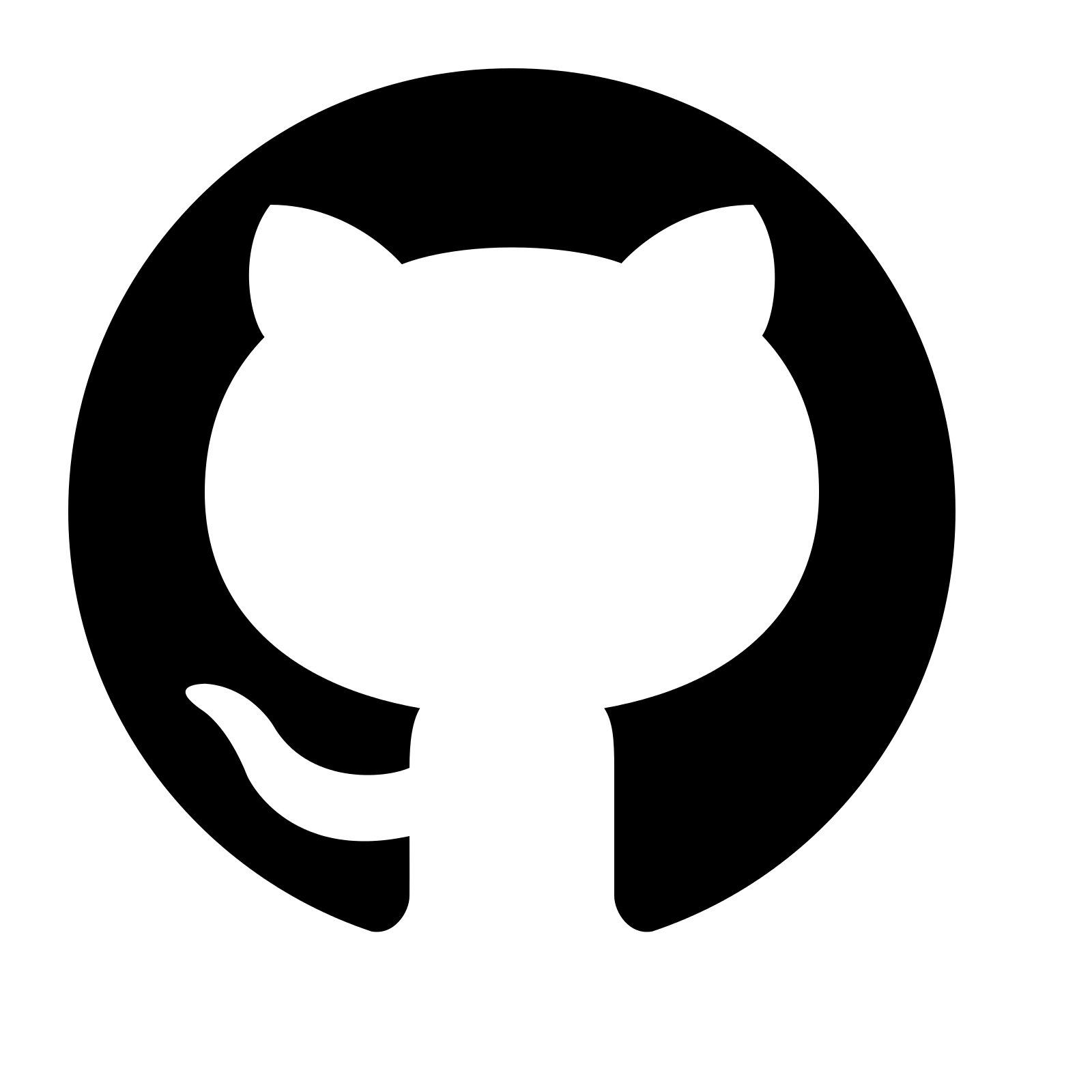}}
        \texttt{ml-jku/neural-gyrokinetics}%
    }
    \hspace{1.0em}
    \href{https://huggingface.co/datasets/gerkone/pinc_gkw}{%
        \raisebox{-0.3em}{\includegraphics[width=1.15em]{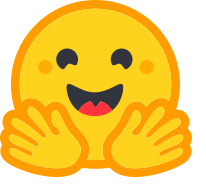}}
        \texttt{gerkone/pinc\_gkw}%
    }
    \vspace{0.5em}
    \end{center}

  \icmlkeywords{Neural Compression, Physics-Informed Neural Networks, Plasma Physics, Gyrokinetics, Turbulence, Neural Fields, Vector Quantized Autoencoder}
]

\printAffiliationsAndNotice{}

\begin{abstract}
High-fidelity scientific simulations are now producing unprecedented amounts of data, creating a storage and analysis bottleneck. A single simulation can generate tremendous data volumes, often forcing researchers to discard valuable information. 
A prime example of this is plasma turbulence described by the gyrokinetic equations: nonlinear, multiscale, and 5D in phase space. 
It constitutes one of the most computationally demanding frontiers of modern science, with runs taking weeks and yielding tens of terabytes of data dumps.
The increasing storage demands underscore the importance of compression. However, reconstructed snapshots do not necessarily preserve essential physical quantities.
We present a spatiotemporal evaluation pipeline, accounting for structural phenomena and multi-scale transient fluctuations to assess the degree of physical fidelity.
Indeed, we find that various compression techniques lack preservation of both spatial mode structure and temporal turbulence characteristics.
Therefore, we explore \ac{PINC}, which incorporates physics-informed losses tailored to gyrokinetics and enables extreme compressions ratios of over $70{,}000\times$.
Entropy coding on top of \ac{PINC} further pushes it to $120{,}000\times$.
This direction provides a viable and scalable solution to the prohibitive storage demands of gyrokinetics, enabling post-hoc analyses that were previously infeasible.
\end{abstract}


\section{Introduction}
Scientific computing is on the cusp of entering an era of high-fidelity simulations across various domains, such as plasma physics \citep{warpx2022, Chang_2024, Dominski2024, kelling2025artificialscientistintransit, disiena2025globalgyrokineticprofilepredictions}, weather and climate modelling \citep{Govett2024exa, bodnar_aurora_2024}, astrophysics \citep{grete2025xmagnetexascalemhdsimulations}, and beyond.
This progress is driven by advancements in High-Performance Computing (HPC): hardware accelerators, exascale computing systems, and scalable numerical solvers are pushing the horizon of what can be computed.
These developments allow practitioners to move beyond reduced numerical approaches and attempt high-fidelity simulations, which are essential to accurately capture the underlying physics of complex systems.
A striking instance of such simulations is gyrokinetics \citep{frieman1982nonlinear, Krommes_gyrokinetics_2012, PEETERS_GKW_2009}, a five-dimensional (5D) nonlinear system that simulates turbulence in magnetised plasmas, such as those found in magnetically-confined fusion devices.

Gyrokinetic simulations generate massive data volumes that create a severe storage and analysis bottleneck.
This arises from their 5D nature, combined with the high-resolution needed to model plasma turbulence.
The gyrokinetic equations express the time evolution of particles in a plasma via a 5D distribution function $\bm{f}\in\mathbb{C}^{v_{\parallel} \times \mu \times s \times x \times y}$, with spatial coordinates $x$, $y$, $s$ and velocity-space coordinates $v_{\parallel}$, $\mu$.
A single run can produce tens of terabytes of data with snapshots saved at many time steps.
In practice, researchers only store diagnostics, making comprehensive post-hoc analysis impossible.
Compression offers a remedy by reducing the cost of storing full 5D fields.
However, no evaluation framework currently exists to assess whether compressed snapshots preserve transient turbulence dynamics, an essential requirement for post-hoc analysis.

As a solution, we introduce an evaluation framework for transient turbulence characteristics in compressed snapshots of gyrokinetic simulations, disentangling \textit{transient fluctuations}, which capture energy transfer across time, from \textit{spatial} quantities that characterize a single snapshot.
Various compression techniques fail to preserve spatial properties, as well as transient turbulence.
To this end, we explore \ac{PINC} for turbulent gyrokinetic data. 
We consider two paradigms: autoencoders (e.g., VQ-VAE \citep{oord2017neural}) generalizing on unseen samples, and neural implicit representations \citep{park2019deepsdflearningcontinuoussigned}, which typically encode individual snapshots into network parameters. 
Unlike conventional compression, \ac{PINC} enforces the preservation of key physical quantities, ensuring that downstream scientific analyses remain valid even at extreme compression rates.

We demonstrate that \ac{PINC} achieves extreme storage reduction while preserving transient turbulence and steady-state spatial characteristics.
Both autoencoders and neural fields attain reconstruction errors comparable to or better than conventional approaches at the same compression rate, while significantly improving physics preservation.
A systematic rate-distortion scaling is also observed between compression rate, signal reconstruction and physics fidelity, allowing this trade-off to be estimated a priori.
Lastly, we showcase additional weight space experiments such as latent temporal interpolation and hybrid compression (\cref{sec:interp}).

We summarize our contributions as follows.
    \circnum{1} A spatiotemporal evaluation pipeline to assess physics preservation. It accounts for both spatial information and temporal dynamics, together capturing multi-scale transient fluctuations prevalent in turbulent dynamic.
    \circnum{2} Physics-Informed, \textit{Gyrokinetics-specific} training curricula for neural compression, \ac{PINC} in short, equipping different techniques with physics losses capturing essential spatial integrals and turbulence characteristics.
    \circnum{3} As a benchmark for future works in (neural) compression, we publish a 50GB gyrokinetics validation set~\footnote{\href{https://huggingface.co/datasets/gerkone/pinc_gkw}{\texttt{gerkone/pinc\_gkw}} on HuggingFace.}, paired with baseline results.

\section{Related Work}

\textbf{Compression} of spatiotemporal data is not a novel topic, and fields such as numerics and HPC conducted a plethora of research in this direction \citep{diffenderfer2019error, isabella2011, lindstrom2014fixed, TTHRESH2019, mo2022pivqae}.
Related research exists in the domain of computational plasma physics \citep{Anirudh_2023}, in particular for Particle-In-Cell (PIC) simulations \citep{Birdsall_PIC_2005, Tskhakaya_PIC_2008}. 
The most relevant works include ISABELA \citep{isabella2011}, an advanced spline method that promises almost lossless compression of spatiotemporal data of up to 7$\times$; and VAPOR \citep{choi2021neural}, a deep learning method based on autoencoders that compresses 2D PIC velocity space slices, supervised with mass, energy and momentum conservation losses.
Concurrent work \citet{kelling2025artificialscientistintransit} proposes streaming pipelines for petascale PIC simulations, learning from data \textit{in-transit} without intermediate storage.
While PIC resolves the full 6D plasma kinetics, gyrokinetics reduces the problem to 5D by averaging over fast gyromotion, enabling turbulent simulations too complex for PIC.
Beyond compression methods, a closely related line of work is super-resolution, which seeks to reconstruct high-resolution fields from coarse inputs \citep{Fukami2023, Yang2025MultiscaleSR, Page2025}.
We address the complementary challenge of compactly storing full snapshots.

\textbf{Implicit Neural Fields} encode information in a compact feature space, enabling scalable, grid-agnostic representation of high-resolution data.
They represent continuous signals as coordinate-based learnable functions \citep{mildenhall2020nerf, park2019deepsdflearningcontinuoussigned, dupont2022functa, mescheder2019occupancynetworkslearning3d}.
In general, neural fields map input coordinates to the respective values of a field, i.e. $f_\theta: \mathbb{R}^d \rightarrow \mathbb{R}^n$~\citep{DBLP:journals/corr/abs-2111-11426}. They are usually implemented as MLPs with special activation functions \citep{sitzmann2020siren, saragadam2023wire, elfwing2017sigmoidweightedlinearunitsneural}. 
In physics, neural fields have been applied to time-varying volumetric data compression~\citep{10.1109/TVCG.2023.3345373} and spatio-temporal dynamics forecasting using implicit frameworks~\citep{DBLP:journals/corr/abs-2306-07266}, among others. 

\textbf{Physics-Informed Neural Networks} (PINNs) combine neural networks with physical constraints \citep{Karniadakis2021}. 
This is typically done by including additional loss terms \citep{raissi2019physics, cai2021physicsinformedneuralnetworkspinns}, ensuring that the laws of physics are obeyed. Constraints such as boundary conditions and conservation laws~\citep{baez2024guaranteeingconservationlawsprojection} are respected in the learned solutions, and more generally that neural network outputs remain consistent with the underlying differential equations. They have been especially effective in solving forward and inverse partial differential equation problems~\citep{raissi2019physics}.
Inversely to the typical local, residual PINN losses, in our case they are global non-linear integrals which depend on the Fourier mode structure.
Sitting at the intersection of PINNs and neural compression, \citet{cranganore2025einsteinfieldsneuralperspective} combine neural fields with Sobolev training \citep{son2021sobolevtrainingphysicsinformed, 10.5555/3294996.3295182} to achieve impressive compression, tensor derivative accuracy and high-fidelity reconstruction on storage intensive general relativity data.
Another notable mention is \citet{mo2022pivqae}, which uses a physics-informed VQ-VAE to capture velocity gradients and statistical properties in isoentropic flows. Our work systematically evaluates whether compressed representations accurately preserve plasma turbulence-specific quantities, motivating the need for specific physics-informed loss terms.

\section{Method}
\subsection{Evaluating Plasma Turbulence}
\label{sec:integrals}

We assess whether compressed representations faithfully capture gyrokinetic turbulence through two complementary groups of metrics, focusing on:
\begin{enumerate*}[label=\textbf{(\arabic*)}]
    \item spatial information, evaluated using non-linear field integrals and turbulence spectra, which measure how well the compressed representations preserve spatial mode structures and energy distributions.
    \item Temporal consistency, via optimal transport of the transitional diagnostics and optical flow distance. These quantify the fidelity of the energy cascade and coherence of the reconstructed sequence.
\end{enumerate*}

\textbfp{Integrals}
In gyrokinetics, (scalar) heat flux $Q\in\mathbb{R}$ and real-space electrostatic potential $\bm{\phi}\in\mathbb{C}^{x\times s \times y}$ are two core quantities.
They describe essential spatial and physical attributes of the density $\bm{f}$.
$Q$ and $\bm{\phi}$ are integrals of the distribution function $\bm{f}$ and are formulated as

\begin{equation}
\label{eq:integrals}
\begin{split}
\bm{\phi} &= \mathbf{A} \int \mathbf{J_{0}} \bm{f} \: \mathrm{d}v_{\parallel}\mathrm{d}\mu, \\
Q &= \int \mathbf{B} \int \mathbf{v}^2 \bm{\phi} \bm{f} \: \mathrm{d}v_{\parallel}\mathrm{d}\mu \:\: \mathrm{d}x\mathrm{d}y\mathrm{d}s.
\end{split}
\end{equation}

$\mathbf{A}, \mathbf{B} \in \mathbb{R}^{x\times s \times y}$ encompass geometric and physical parameters, 
$\mathbf{v} \in \mathbb{R}^{v_{\parallel} \times \mu}$ is the particle energy,
and $\mathbf{J_{0}}$ denotes the 0th-order Bessel. The electrostatic potential $\bm{\phi}$ is obtained by integrating the velocity-space of $\bm{f}$, while the heat flux $Q$ depends on both $\bm{f}$ and $\bm{\phi}$.
Intuitively, $\bm{\phi}$ represents spatial variations of the electric field, while $Q$ is the energy flow carried by particles along field lines.

\textbfp{Wavespace distribution (diagnostics)}
These derived quantities are used by researchers to determine the properties of a simulation and to diagnose the soundness of a given configuration; they measure how energy and electrostatic fluctuations are distributed across modes in wavenumber space, and provide a basis for identifying behaviors and patterns in turbulent plasma transport.
In particular, $k_y^{\text{spec}}\in\mathbb{C}^{N_{k_y}}$ describes the perpendicular scales of turbulence along $y$, and $Q^{\text{spec}}\in\mathbb{C}^{N_{k_y}}$ links turbulent structures to heat transport. They are expressed as convolutions of $\bm{\phi}$ and $\bm{Q}$,

\begin{equation}
\label{eq:diag}
\begin{split}
k_y^{\text{spec}}(y) &= \sum_{s, x} |\hat{\bm{\phi}}(x, s, y)|^2 \:,\\
Q^{\text{spec}}(y) &= \sum_{v_{\parallel}, \mu, s, x} \bm{Q}(v_{\parallel}, \mu, s, x, y)  \:,
\end{split}
\end{equation}

where $\hat{\bm{\phi}}$ is the Fourier space electrostatic potential, and $\bm{Q}$ is the heat flux field (also in Fourier space) before applying the outermost integral, which aggregates it to $Q$.
Diagnostics are used to characterize turbulence, and can be analyzed both in a time-averaged or transient manner.
Time dependency is used to observe how the energy cascade shifts in the energy to lower modes and vice versa, while statistically-steady forms (time-averaged, $\overline{k_y^{\text{spec}}}$ and $\overline{Q^{\text{spec}}}$) define dominant modes. Namely, $\overline{k_y^{\text{spec}}}$ is the mean turbulent spectrum, and $\overline{Q^{\text{spec}}}$ quantifies the heat flux contribution to turbulent transport. They are both used by researchers to detect whether turbulence develops and at which scale.

\textbfp{Time dynamics}
Turbulence is inherently a spatiotemporal phenomena, and a purely spatial evaluation is insufficient to assess reliable reconstruction. To that end, we include metrics from two different perspectives to quantify temporal consistency.
First, the fidelity at which the onset of turbulence is reproduced can be assessed in the transitional phase, between the linear and the statistically-steady state.
We quantitatively evaluate the time-accumulated optimal transport of the wavespace distributions $k_y^{\text{spec}}$ and $Q^{\text{spec}}$ (\cref{eq:diag}) through Wasserstein Distance (WD), and denote it as Energy Cascade (EC) error.
It captures how well the bi-directional energy cascade is captured.
Given two diagnostics sequences of $N$ timesteps in the transition phase, $k_y^{\text{spec}},Q^{\text{spec}}$ and predicted $\widehat{k}_y^{\text{spec}},\widehat{Q}^{\text{spec}}$,
\begin{equation}
\label{eq:energy_cascade}
\begin{split}
\mathrm{EC}_{k_y} &= \sum_{i=1}^N \text{WD}(k_{y,i}^{\text{spec}}, \widehat{k}_{y,i}^{\text{spec}}), \\
\mathrm{EC}_{Q} &= \sum_{i=1}^N \text{WD}(Q_{i}^{\text{spec}}, \widehat{Q}_{i}^{\text{spec}}).
\end{split}
\end{equation}

Second, to check the dynamic consistency of the decompressed sequence we employ the \textit{EndPoint Error} (EPE) of the optical flow field \citep{Baker2011epe}, commonly used in video modeling \citep{Argaw2022LongtermVF,ma2024vidpanos}. 
Additional definition in \cref{app:temporal_metrics}.
Given two sequences of $x_1$ and $x_2$ of $N$ frames and their $i$-th flow vectors $\mathbf{F}_1^{(i)}$ and $\mathbf{F}_2^{(i)}$, the EndPoint Error is
\begin{equation}
\label{eq:endpoint}
\mathrm{EPE}(x_1, x_2) = \frac{1}{N} \sum_{i=1}^N \|\mathbf{F}_1^{(i)} - \mathbf{F}_2^{(i)} \|_2.
\end{equation}

\subsection{Neural Compression}

\begin{figure}[t]
\centering
\includegraphics[width=0.9\linewidth, trim=50px 0 50px 0, clip]{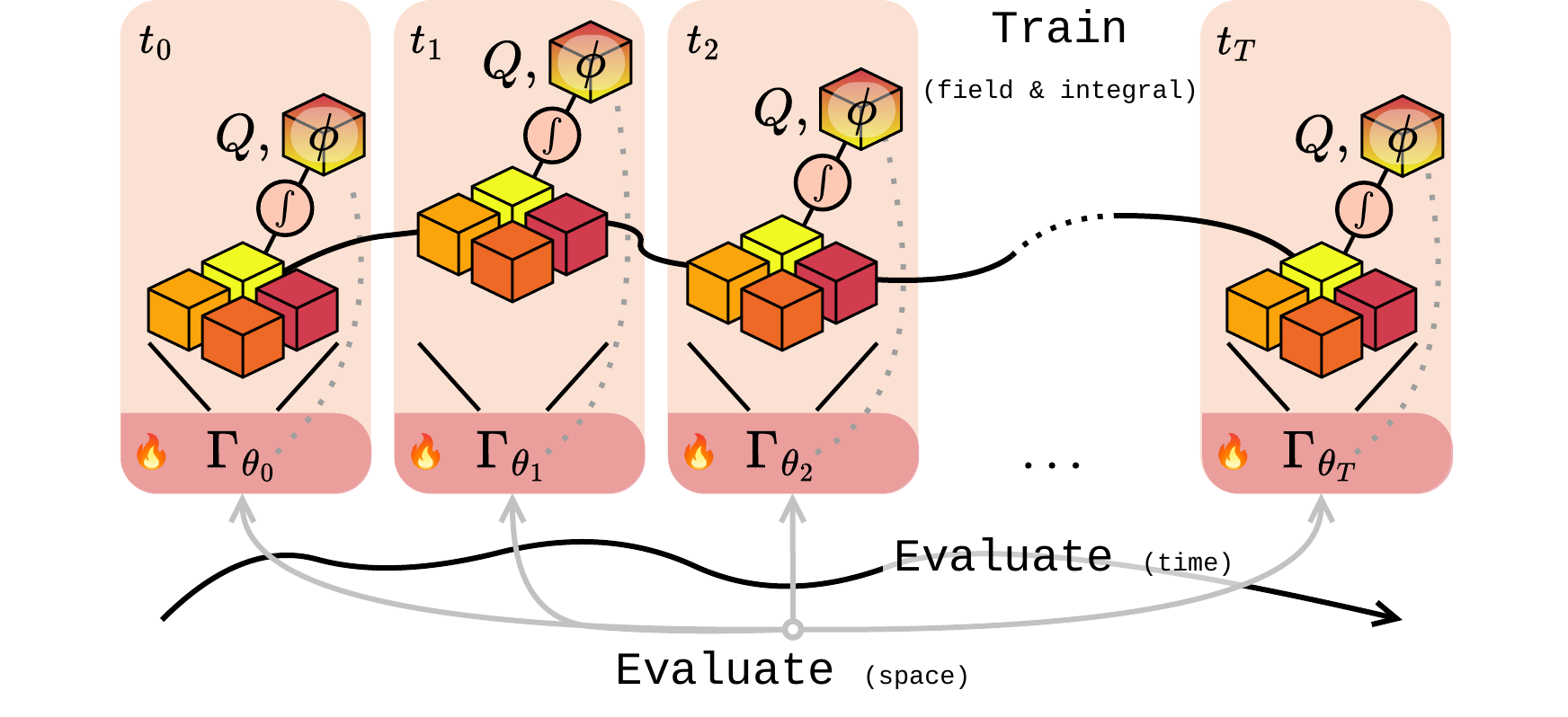}
\caption{
Sketch of the training and evaluation for \acf{PINC} models. 
Training is done at individual time snapshots, while evaluation considers turbulence taking both spatial and temporal information into account.
}
\label{fig:figure1}
\end{figure}

We identify two dominant approaches to learned compression, depending on a few key aspects. 
The first approach are autoencoders, with explicit latent space compression at the bottleneck between an encoder and a decoder.
Parameters $\theta$ are shared across snapshots and time, and a single model $\Gamma_{\theta}$ is trained on a dataset. Compression is applied to unseen samples.
VQ-VAE \citep{oord2017neural} exemplifies autoencoders designed for compression. 
In contrast, neural implicit representations overfit an independent set of parameters at each datapoint, for instance across time $[\Gamma_{\theta_t}]_{(0\dots T)}$.
Encoding is implicit in weight-space and reconstruction happens by querying the neural field.
\cref{fig:figure1} outlines \ac{PINC} training and evaluation for a trajectory.
%
The complex Mean Squared Error (cMSE) on the density $\bm{f}$ is used as reconstruction loss in training
\begin{equation}
\label{eq:recon_loss}
\mathcal{L}_{\text{recon}} = \sum_{v_{\parallel},\mu,x,s,y} \left[ \Re(\bm{f}_{\text{pred}} - \bm{f}_{\text{GT}})^2 + \Im(\bm{f}_{\text{pred}} - \bm{f}_{\text{GT}})^2 \right] \:.
\end{equation}

\textbfp{5D autoencoders}
\label{sec:ae}
Due to the high-dimensional nature of the data, we leverage nD swin layers \citep{galletti20255dneuralsurrogatesnonlinear,paischer2025gyroswin}, based on Shifted Window Attention \citep{liu_swin_2021}, which promise scaling to higher dimensions. They work by first partitioning the domain in non-overlapping \textit{windows}, then performing attention only locally within the window.
An autoencoder $\Gamma_{\theta}: \mathbb{C}^{(v_{\parallel}, \mu, s, x, y)} \times \mathbb{R}^4 \to \mathbb{C}^{v_{\parallel}, \mu, s, x, y}$, with $\Gamma_{\theta}(\bm{f}, \bm{c}) = \mathcal{D} \circ \mathcal{E}(\bm{f}, \bm{c})$, encodes the 5D density field $\bm{f}$ and conditioning $\bm{c}$ containing four gyrokinetic parameters ($R/L_T$, $R/L_n$, $q$, and $\hat{s}$) into a compact latent space, then decodes it to reconstruct $\bm{f}$.
Following hierarchical vision transformers, the encoder $\mathcal{E}$ tiles $\bm{f}$ into patches and applies interleaved Swin and downsampling layers. At the bottleneck, the latent dimension is downprojected to control the compression rate.
The decoder $\mathcal{D}$ mirrors this, with upsampling to restore the original resolution.
We apply both regular Autoencoders (AE) and Vector-Quantized Variational Autoencoders (VQ-VAEs) \citep{oord2017neural}.
Autoencoders are \textit{monolithic models} that compress in an explicit latent space, enabling cheap compression and decompression. However, they usually require expensive pretraining and a diverse dataset to generalize across different simulations.

\textbfp{Neural implicit fields}
The distribution function $\bm{f}$ is indexed by a five-tuple of coordinates $(v_{\parallel}, \mu, s, x, y)$. Specifically, we train a separate (discrete) coordinate-based Neural Field $\Gamma_{\theta_{t, c}}: \mathbb{N}^5 \to \mathbb{C}$ to fit each $\bm{f}_t^{c}$ at time $t$ of a trajectory configured by $c$.
Indices are encoded with a learnable embedding hashmap \citep{mueller2022instant}, then passed to an MLP using SiLU \citep{elfwing2017sigmoidweightedlinearunitsneural}, sine \citep{sitzmann2020siren} or Gabor \citep{saragadam2023wire} activations.
Fitting a $\Gamma_{\theta_{t, c}}$ takes $\sim$1-2 minutes (NVIDIA H100). Since independent networks are used per snapshot, training is fully parallelized and performed in a staggered / pipelined fashion.
Neural fields are \textit{micromodels}: individual samples are implicitly encoded into network weights, offering resolution invariance and lower training requirements. Conversely, encoding is relatively costly.

\subsection{\acf{PINC}}
\label{sec:pinn}
Training on $\mathcal{L}_{\text{recon}}$ alone cannot ensure conservation of physical quantities or turbulent characteristics. Further, due to the limited representation power, lossy compression inevitably discards useful information if left unconstrained.
We supervise on the physical quantities listed in \cref{sec:integrals} by penalizing (absolute) deviations from the ground truth. 
Integral and wavespace losses are defined as
\begin{equation}
\label{eq:int_loss}
\begin{aligned}
\colorbox{green!10}{$\mathcal{L}_Q$} &= | Q_\text{pred} - Q_\text{GT} |, 
& \colorbox{green!10}{$\mathcal{L}_{\bm{\phi}}$} &= \text{L1}(\bm{\phi}_{\text{pred}}, \bm{\phi}_{\text{GT}}), \\
\colorbox{yellow!10}{$\mathcal{L}_{k_y}$} &= \text{L1}(k_{y\text{, pred}}^{\text{spec}}, k_{y\text{, GT}}^{\text{spec}}),
& \colorbox{yellow!10}{$\mathcal{L}_{Q^{\text{spec}}}$} &= \text{L1}(Q^{\text{spec}}_{\text{pred}}, Q^{\text{spec}}_{\text{GT}}).
\end{aligned}
\end{equation}
In addition, we introduce a first-order constraint to capture the turbulent energy cascade.
In the case of simulations with a single energy injection scale, the spectra must be monotonically decreasing after the dominant mode, indexed by the peak wavenumber $k_{\text{peak}}$.
This specific monotonicity loss can be written as the log-transformed isotonic loss, penalizing negative slopes,
\vspace{-6px}
\begin{equation}
\label{eq:mono_loss}
\colorbox{blue!10}{$\mathcal{L}_{\text{iso}}(\mathbf{s})$} =
\frac{1}{N - k_{peak}} \sum_{i=k_\text{peak}}^{N-1} 
\Big| \textrm{ln} [ \mathbf{s}(i) ] - \textrm{ln} [ \mathbf{s}(i)^\text{sorted} ] \Big|.
\end{equation}
With $\mathbf{s}\in\mathbb{C}^N$ the diagnostic vector ($k_y^{\text{spec}}$ and $Q^{\text{spec}}$).
Combining all terms yields the final physics-informed loss:
\begin{equation}
\label{eq:final_loss}
\begin{split}
\mathcal{L}_{\text{PINC}} = & \left. \colorbox{green!10}{$\mathcal{L}_Q + \mathcal{L}_{\bm{\phi}}$} \right\}_{\mathcal{L}_{\text{int}}} + \\
& \left. \colorbox{yellow!10}{$\mathcal{L}_{k_y^{\text{spec}}} + \mathcal{L}_{Q^{\text{spec}}}$} \right\}_{\mathcal{L}_{\text{diag}}} + \\
& \left. \colorbox{blue!10}{$\mathcal{L}_{\text{iso}}(k_{y\text{, pred}}^{spec}) + \mathcal{L}_{\text{iso}}(Q^{spec}_{\text{pred}})$} \right\}_{\mathcal{L}_{\text{grad}}} \,.
\end{split}
\end{equation}

\begin{table*}[b]
\renewcommand{\arraystretch}{1.0}
\scriptsize
\caption{Comparison between \ac{PINC} and traditional methods.
Evaluation on 420 total $\bm{f}_t^c$s (20 turbulent trajectories, 21 timesteps), sampled in the statistically steady phase.
Errors in data space. Best result in bold, second best underlined.
\label{tab:results}}
\centering
\resizebox{\textwidth}{!}{%
\begin{tabular}{lr|cc|cc|cc}
\toprule
          &       & \multicolumn{2}{c}{Compression $\bm{f}$}                               & \multicolumn{2}{c}{Integrals $Q,\bm{\phi}$}                                               & \multicolumn{2}{c}{Turbulence $Q^{\text{spec}}, k_y^{\text{spec}}$}                                         \\ \midrule
          & CR 
          & PSNR $\uparrow$  & $\text{EPE}$ $\downarrow$  &
          $\text{L1}(Q)$ $\downarrow$ & $\text{PSNR}(\bm{\phi})$ $\uparrow$ &
          $\text{WD}(\overline{k_y^{\text{spec}}})$ $\downarrow$ & $\text{WD}(\overline{Q^{\text{spec}}})$ $\downarrow$ \\\midrule
ZFP      & 901$\times$ & 28.97$_{\pm 1.09}$ & 0.175$_{\pm 0.07}$ & 107.48$_{\pm 49.35}$ & -16.20$_{\pm 7.09}$ & 0.020$_{\pm 0.01}$ & 0.116$_{\pm 0.20}$ \\
Wavelet  & 936$\times$ & 33.07$_{\pm 1.18}$ & 0.064$_{\pm 0.03}$ & 107.74$_{\pm 49.51}$ & -13.24$_{\pm 9.20}$ & 0.020$_{\pm 0.01}$ & \textbf{0.010$_{\pm 0.00}$} \\
PCA      & 1020$\times$ & 32.09$_{\pm 0.98}$ & 0.123$_{\pm 0.07}$ & 67.60$_{\pm 36.08}$ & -10.22$_{\pm 6.89}$ & 0.020$_{\pm 0.01}$ & 0.011$_{\pm 0.00}$ \\
JPEG2000 & 1000$\times$ & 34.33$_{\pm 0.95}$ & 0.046$_{\pm 0.02}$ & 103.91$_{\pm 44.12}$ & -20.85$_{\pm 6.50}$ & 0.020$_{\pm 0.01}$ & 0.035$_{\pm 0.03}$ \\
\midrule
NF       & 1167$\times$ & \textbf{36.91$_{\pm 0.93}$} & \textbf{0.031$_{\pm 0.02}$} & 61.50$_{\pm 16.91}$ & 1.24$_{\pm 5.99}$ & 0.017$_{\pm 0.01}$ & 0.017$_{\pm 0.00}$ \\
PINC-NF  & 1167$\times$ & \underline{35.76$_{\pm 1.38}$} & \underline{0.037$_{\pm 0.02}$} & \textbf{2.18$_{\pm 8.33}$} & \textbf{13.50$_{\pm 4.44}$} & \textbf{0.006$_{\pm 0.00}$} & 0.015$_{\pm 0.00}$ \\
AE + EVA & 716$\times$ & 35.64$_{\pm 2.03}$ & 0.063$_{\pm 0.05}$ & \underline{15.01$_{\pm 16.42}$} & 6.72$_{\pm 4.98}$ & 0.016$_{\pm 0.01}$ & 0.012$_{\pm 0.01}$ \\
VQ-VAE + EVA & 77368$\times$ & 32.61$_{\pm 1.58}$ & 0.095$_{\pm 0.07}$ & 44.26$_{\pm 40.97}$ & \underline{7.66$_{\pm 3.75}$} & \underline{0.015$_{\pm 0.01}$} & 0.013$_{\pm 0.01}$ \\
\bottomrule
\end{tabular}
}
\end{table*}

Importantly, our training supervises the model on \emph{nonlinear integrals} of the distribution function, rather than directly on PDE residuals \citep{Karniadakis2021} or derivatives \citep{son2021sobolevtrainingphysicsinformed}. 
\ac{PINC} implicitly directs the network to the physically relevant modes.
In turn, as $Q$ and $\bm{\phi}$ integrals depend on the full spectral structure of $\bm{f}$, many of the losses in \cref{eq:final_loss} are \textit{global} quantities, rather than the local pointwise supervision typical in PINNs.
$\mathcal{L}_{\text{PINC}}$ can be included in training, but with two caveats:
\begin{enumerate*}[label=(\textbf{\roman*})]
    \item loss terms depend on $\bm{f}$'s mode composition, and
    \item global loss terms cannot be computed at coordinate-level.
\end{enumerate*}
We address (\textbf{i}) by applying $\mathcal{L}_{\text{PINC}}$ after $\bm{f}$'s have converged, to ensure that structure is present.
(\textbf{ii}) is problematic only for local or sparse methods. The following sections detail the devices enabling \ac{PINC} training on neural fields and autoencoders.

\textbfp{\ac{PINC}-neural fields}
Neural fields fit $\mathcal{L}_{\text{PINC}}$, continuing optimization after the initial steps where $\bm{f}$ is learned.
For stability, multi-objective optimizers offer a more principled alternative to manual tweaking or schedulers.
Conflict-Free Inverse Gradients \citep[ConFIG]{Liu2024ConFIG} and Augmented Lagrangian Multipliers \citep{basir2023adaptiveaugmentedlagrangianmethod} are sometimes employed in tasks with many competing losses \citep{berzins2025geometryinformedneuralnetworks}.
Finally, even though neural fields are normally trained on small sparse coordinate batches, $\mathcal{L}_{\text{PINC}}$ gradients can only be computed on the entire grid.

\textbfp{\ac{PINC}-autoencoders}
Training autoencoders with physics constraints across heterogeneous samples tends to result in training instabilities. To mitigate this, we employ parameter-efficient finetuning.
Specifically, we pre-train the autoencoder on $\mathcal{L}_{\text{recon}}$, and finetune it on $\mathcal{L}_{\text{PINC}}$ using Explained Variance Adaptation \citep[EVA]{paischer2025eva}, an improved and more stable variant of LoRA-style adapters \citep{hu2022lowrank}.
More training details in \cref{app:swin}.

\section{Results}
Neural fields are simple MLPs with SiLU activations \citep{elfwing2017sigmoidweightedlinearunitsneural}, 64 latent dimension, 5 layers and skip connections.
They are fit using AdamW \citep{loshchilov2019decoupledweightdecayregularization} with learning rate decaying between $[5e-3, 1e-12]$ (details in \cref{app:nf}).
Results suggest that neural fields trained with ConFIG are less accurate on physical losses, but lead to a marginally better reconstruction error (Appendix \cref{tab:config_results}).
For simplicity, all models reported are trained without loss balancing, unless specified otherwise.
Grid searches and ablations are in \cref{app:nf_ablations}.

As for standard autoencoders and VQ-VAEs,
swin tokens are 1024-dimensional, bottleneck dimension is 32, and the codebook dimension of the VQ-VAE is 128, totaling at $\sim$152M parameters.
Both are trained and fine-tuned on 6,890 $\bm{f}$ time snapshots, amounting to around 500GB of data (details in \cref{app:dataset}).
Compression and reconstruction is subsequently expected to happen \textit{out of distribution}, to unseen trajectories.
Pre-training takes $\sim$60 hours (200 epochs, 4$\times$ NVIDIA H100) for standard AE and VQ-VAE. Fine-tuning with EVA weights takes $\sim$28 hours on one NVIDIA H100 for 120 epochs, adapting $\sim$4\% (6M) of the total parameters. Optimized using Muon \citep{jordan2024muon} with cosine scheduling of the learning rate between $[2e-4, 4e-6]$ (details in \cref{app:swin}).

We compare with traditional compression based on different techniques: ZFP \citep{lindstrom2014fixed}, a very popular compression method for scientific data relying on block-quantization, Wavelet-based compression, spatial PCA and JPEG2000 adapted for the 5D data. Baselines are tuned to achieve compression rates (CRs) of around $1{,}000\times$ (99.9\% size reduction), comparable with neural fields and vanilla autoencoders. For reference, \textit{off-the-shelf} traditional techniques such as \texttt{gzip} achieve a lossless compression ratio of $\sim$1.1x (8\% reduction). Information on baselines can be found in \cref{app:trad}.
Detailed information about runtime can be found in the Appendix \cref{tab:timing}.

\subsection{Compression}
\label{sec:compress}
\cref{tab:results} quantitatively summarizes the compression and reconstruction results.
We evaluate all methods on traditional compression metrics, integral, and turbulence errors.
To measure spatial $\bm{f}$ reconstruction quality after compression, Peak Signal-to-Noise-Ratio (PSNR) is reported (defined in \cref{app:metrics}).
To evaluate temporal compression, we report the EndPoint Error (EPE) (\cref{eq:endpoint}) for turbulent snapshots of $\bm{f}$.
Integral errors are reported as mean absolute error of flux $Q$ and potential $\bm{\phi}$ after integration of $\bm{f}$ according to \cref{eq:integrals}. 
For steady-state turbulence evaluation we normalize the \textit{time-averaged}, $\overline{k_y^{\text{spec}}}$ and $\overline{Q^{\text{spec}}}$ spectra and employ Wasserstein Distance (WD), which is commonly used as a geometry-aware distance metric and can efficiently be computed for 1D spectra. 
We report additional metrics for spatial evaluation in \cref{tab:missing_results}.
Furthermore, for discussion on transient dynamics see Paragraph \ref{sec:turb} (\cref{fig:cascade}).
Qualitative examples of reconstructions for $\bm{f}$ and $\bm{\phi}$ are in the Appendix at Figures \ref{fig:extra_df} and \ref{fig:extra_phi}.

At similar compression, non-\ac{PINC} neural fields and autoencoders improve upon traditional methods on quality, as well as integrated quantities and turbulence.
However, especially integral metrics ($\text{L1}(Q)$ and $\text{PSNR}(\bm{\phi})$) exhibit excessive discrepancies from the ground-truth.
This motivates \ac{PINC} soft-constraints on the optimization to preserve them.

Comparing NF to \ac{PINC}-NF reveals improvements on integral errors at a minor reconstruction degradation.  
Furthermore, WD decreases by almost $3\times$ for $\overline{k_y^{\text{spec}}}$.
Interestingly, we do not observe an improvement on $\overline{Q^{\text{spec}}}$, possibly due to competing objectives.
A possible interpretation for these two patterns is that, 
since modeling capacity is constrained by high compression, the available ``entropy'' gets allocated across modes, according to the encoding algorithm. In neural networks, the spectral bias \citep{rahaman2019spectralbiasneuralnetworks} of MSE training (\cref{eq:recon_loss}) implies that high-energy components have priority during training, while lower modes converge slower.
\ac{PINC} appears to redistribute some of the energy to more physically relevant modes. For example, the heat flux integral masks low frequencies and rescales high frequencies, giving them more importance.

\begin{figure}[!t]
\centering
    \includegraphics[height=0.62\linewidth]{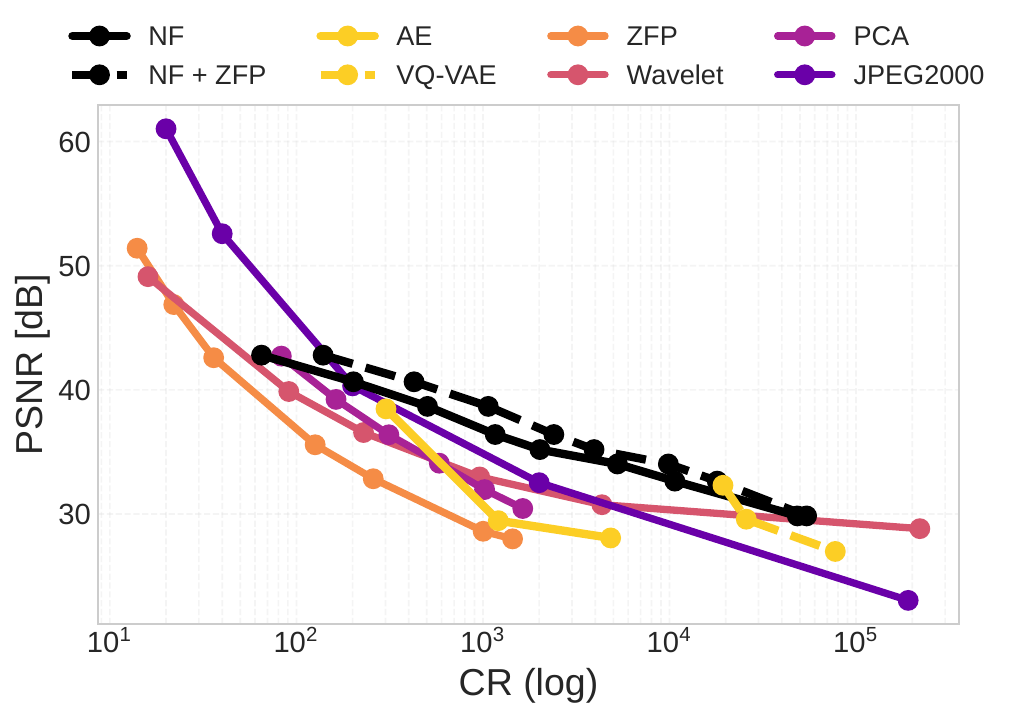}
\caption{
Compression performance rate-distortion as Peak Signal to Noise Ratio (PSNR) on Compression Rate (CR) on 3 randomly sampled timesteps from 10 trajectories (30 total samples).
}
\label{fig:scaling}
\end{figure}

\textbfp{Performance-rate scaling}
To assess how reconstruction quality scales across compression levels, we train a series of neural fields and autoencoders with progressively larger parameter counts and latent sizes. Training neural fields remains relatively inexpensive, whereas autoencoders become unfeasible in terms of both GPU memory and runtime at lower CRs. Consequently, we train only six autoencoders in total (three standard and three VQ-VAEs), all at comparatively high CRs (>$1{,}000\times$).
Findings reported in \cref{fig:scaling} suggest that both learned methods present a specific "window" of CRs in which they significantly outperform traditional baselines (namely in the $500-10{,}000\times$ range).
Moreover, neural fields also exhibit a favorable exponential decay (linearly in semilog-x), as opposed to super-exponential of others (polynomial in semilog-x). This is supported by neural field compression on other modalities \citep{dupont2022coinneuralcompressionmodalities, spatialfuncta23}.
In terms of reconstruction quality, at lower rates ($<200\times$) neural compression cannot reliably match wavelets or JPEG2000, and at extreme CRs ($>40{,}000\times$) they are comparable.

\subsection{Physics and Turbulence Preservation}
\begin{figure}[!t]
\centering
\includegraphics[height=0.62\linewidth]{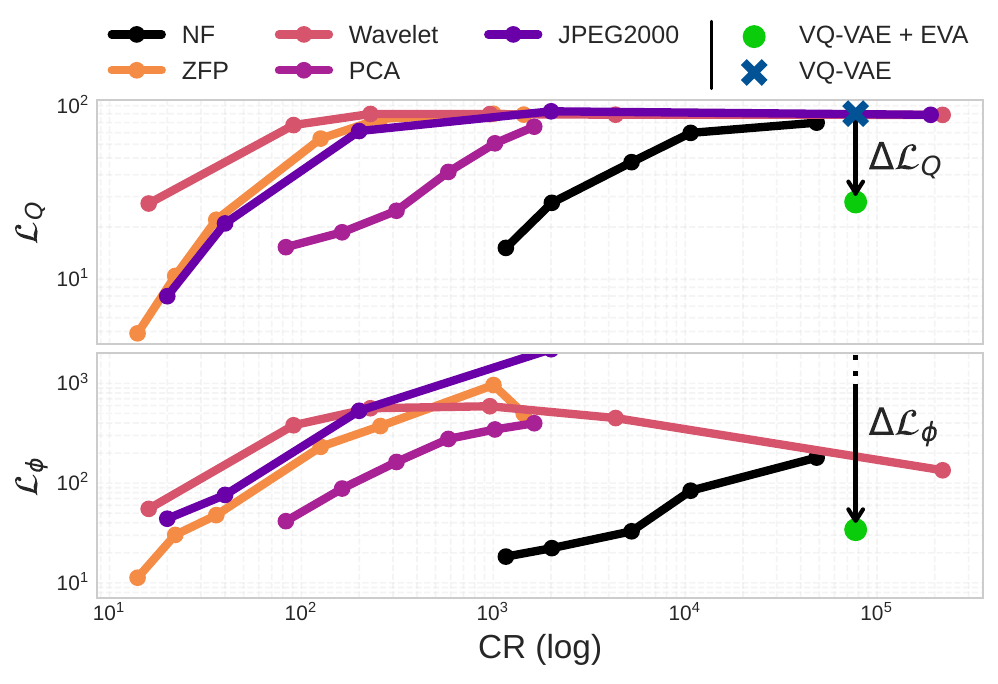}
\caption{
Physical loss scaling as $\mathcal{L}_{Q}$ (top) and $\mathcal{L}_{\bm{\phi}}$ (bottom) on Compression Rate (log-log). NF and VQ-VAE + EVA are trained with \ac{PINC} losses. Arrow denotes $\Delta\mathcal{L}$ improvement for VQ-VAE.
}
\label{fig:phisics_scaling}
\end{figure}

\textbfp{Physical scaling}
\label{sec:pinn_scaling}
Similarly to \cref{fig:scaling} for rate-distortion for $\bm{f}$, \cref{fig:phisics_scaling} shows scaling for heat flux $Q$ and electrostatic potential $\bm{\phi}$ integral losses as CR is changed. \cref{fig:extra_phi} shows corresponding projections of the 3D $\bm{\phi}$ integral and residuals ($\text{CR}=\sim1{,}000\times$). Traditional compression struggles to capture $\phi$ even at low CR, while \ac{PINC} models trained on \cref{eq:final_loss} as well as the reconstruction loss (\cref{eq:recon_loss}) yield reasonable reconstruction.

\begin{figure}[!b]
\centering
\vspace{-8px}
\includegraphics[height=0.61\linewidth]{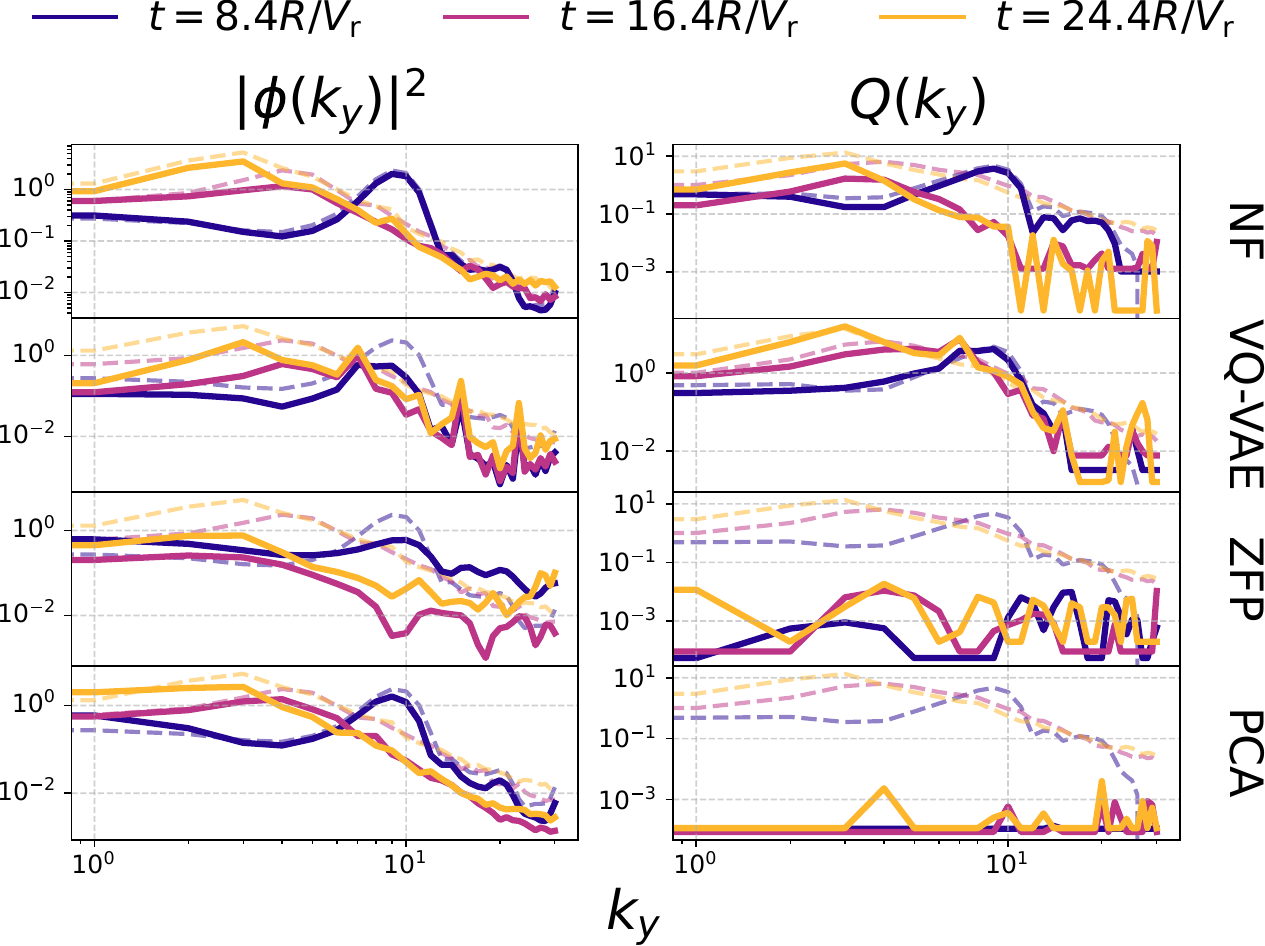}
\caption{
Energy cascade visualized as the transfer from higher to lower modes as turbulence develops. Plots in loglog scale.
}
\label{fig:cascade}
\end{figure}
\textbfp{Recovering turbulence}
\label{sec:turb}
\cref{fig:cascade} qualitatively shows how well different models capture the direct energy cascade phenomena across different simulations (energy shifting to lower modes over time), by visualizing the \textit{per-timestep} spectras $k_y^{\text{spec}}$ and $Q^{\text{spec}}$ in a log-log plot. The Figure provides a qualitative comparison of turbulence recovery on the temporal axis, in contrast to the time-averaged statistics reported in \cref{tab:results}.
The time snapshots examined in \cref{fig:cascade} (sampled between $[8.4, 24.4]R/V_{\mathrm{r}}$ with a step size of $\Delta=2.0R/V_{\mathrm{r}}$) are sampled in the transitional phase where turbulence grows, at the so called \textit{overshoot}. 
These timesteps are different to those in \cref{tab:results}.
On $k_y^{\mathrm{\text{spec}}}$, traditional compression methods already achieve reasonable performance in most cases, but on $Q^{\text{spec}}$ they produce severely nonphysical results (flat curves, negative numbers). Another observation is that, even though non-ML methods have fairly low Wasserstein distance in \cref{tab:results}, this is not reflected at the overshoot.
In contrast, neural fields and VQ-VAE can reproduce the overall profiles consistently, with VQ-VAE excelling at the flux spectra. However, both often fail to capture the higher-frequency magnitudes. The behaviors can be attributed to the spectral bias of neural networks \citep{rahaman2019spectralbiasneuralnetworks, teney2025neuralredshiftrandomnetworks}, where low-frequency (high-energy) components are favored over high-frequencies.
\cref{app:extra_results} shows additional cascade plots for all methods and trajectories.

\begin{table}[t]
    \centering
    \scriptsize
    \caption{
    Temporal consistency: endpoint error (EPE) and energy cascade Wasserstein (EC). Evaluation on 270 $\bm{f}_t^c$s (30 trajectories, 9 timesteps) from the transition phase where mode growth happens.
    \label{tab:temporal}}
    \vspace{-10px}
    \renewcommand{\arraystretch}{1.1}
    \setlength{\tabcolsep}{4pt}
    \resizebox{0.95\linewidth}{!}{%
    \begin{tabular}{lccc}
        \toprule
        Model & EPE $\downarrow$ & $\text{EC}_{k_y}$ $\downarrow$ & $\text{EC}_{Q}$ $\downarrow$ \\
        \midrule
        ZFP              & 0.058$_{\pm 0.03}$ & 0.031$_{\pm 0.01}$ & 0.715$_{\pm 1.30}$ \\
        Wavelet          & 0.033$_{\pm 0.01}$ & 0.031$_{\pm 0.01}$ & 0.061$_{\pm 0.09}$ \\
        PCA              & 0.032$_{\pm 0.02}$ & 0.032$_{\pm 0.01}$ & 0.065$_{\pm 0.07}$ \\
        JPEG2000         & \uline{0.027}$_{\pm 0.01}$ & 0.032$_{\pm 0.01}$ & 0.176$_{\pm 0.21}$ \\ \midrule
        NF               & \textbf{0.017}$_{\pm 0.01}$ & 0.030$_{\pm 0.01}$ & 0.029$_{\pm 0.02}$ \\
        \ac{PINC}-NF     & 0.030$_{\pm 0.02}$ & \textbf{0.011}$_{\pm 0.01}$ & 0.015$_{\pm 0.00}$ \\
        \ac{PINC}-AE     & 0.030$_{\pm 0.02}$ & 0.028$_{\pm 0.01}$ & \textbf{0.005}$_{\pm 0.00}$ \\
        \ac{PINC}-VQ-VAE & 0.036$_{\pm 0.02}$ & \uline{0.018}$_{\pm 0.01}$ & \uline{0.008}$_{\pm 0.00}$ \\
        \bottomrule
    \end{tabular}
    }
\vspace{-8px}
\end{table}

\cref{tab:temporal} shows that neural compression can significantly outperform traditional methods both on energy cascade errors (\cref{eq:energy_cascade}), as well as the endpoint error of $\bm{f}$'s optical flow (\cref{eq:endpoint}).
We observe that \ac{PINC}-NFs get less aligned over time, as reflected by the higher EPE and motivated to the performance drop discussed in \cref{sec:compress}. Conversely, they significantly outperform unregularized NFs on the two energy cascade metrics, suggesting a better representation of turbulence. 
Note that the EPE reported here differs from the one in \cref{tab:results}: here it is applied to the transition phase, instead of the statistically steady one.

\textbfp{Physical losses ablations}
\begin{table}[!t]
\centering
\scriptsize
\renewcommand{\arraystretch}{1.1}
\setlength{\tabcolsep}{2pt}
\caption{
PINC losses ablations from \cref{eq:final_loss} (color-coded blocks). * means $>100\times$ larger than column average.
\label{tab:pinn_ablations}}
\resizebox{0.95\linewidth}{!}{%
\begin{tabular}{llccccc}
    \toprule 
    Model & \multicolumn{1}{c}{Loss} & PSNR($\bm{f}$) & $\mathcal{L}_{Q}$ & $\mathcal{L}_{\bm{\phi}}$ & $\mathcal{L}_{ky^{\text{spec}}}$ & $\mathcal{L}_{Q^{\text{spec}}}$ \\ 
    \midrule 
    \multirow{5}{*}{NF}
    & $\phantom{+}\mathcal{L}_{\text{recon}}$ & \textbf{38.89} & 48.59 & 4.45 & 3.71 & 1.52 \\
    & \cellcolor{green!10}{$+ \mathcal{L}_{\text{int}}$} & 36.68 & \textbf{10.35} & 2.55 & 1.61 & 1.42 \\
    & \cellcolor{yellow!20}{$+ \mathcal{L}_{\text{diag}}$} & 38.76 & 41.39 & 2.25 & 1.67 & \textbf{1.32} \\
    & \cellcolor{blue!10}{$+ \mathcal{L}_{\text{grad}}$} & 37.29 & 63.94 & 44.18 & * & 2.0 \\
    & $+ \mathcal{L}_{\text{PINC}}$ & 38.28 & 28.03 & \textbf{1.41} & \textbf{0.24} & 1.41 \\
    \midrule 
    \multirow{2}{*}{VQ-VAE}
    & $\phantom{+}\mathcal{L}_{\text{recon}}$ & 26.96 & 86.21 & * & * & 91.68 \\
    & $+ \mathcal{L}_{\text{PINC}}$           & 27.73 & 85.06 & 103.50 & * & * \\
    + EVA & $+ \mathcal{L}_{\text{PINC}}$     & \textbf{32.21} & \textbf{27.73} & \textbf{40.81}  & \textbf{284.96} & \textbf{59.84} \\
    \bottomrule
\end{tabular}
}
\vspace{-8px}
\end{table}
We verify the impact of each loss term described in \cref{eq:final_loss} by training different models on each term in \cref{sec:integrals} and \cref{sec:pinn} separately, for both autoencoders and neural fields.
\cref{tab:pinn_ablations} collects the ablation findings.
Training $\mathcal{L}_{\text{int}}$ and $\mathcal{L}_{\text{diag}}$ have similar effects, both improve the integral as well as the diagnostics, with the integral being more informative.
The model still gets valuable information on $Q$ and $\bm{\phi}$ from the gradients through $\mathcal{L}_{\text{diag}}$.
In contrast, $\mathcal{L}_{\text{grad}}$ alone has a destabilizing effect, and is only effective when combined with other losses as it is dependent on how accurately the diagnostics (and integrals) are captured.
Finally, the composite $\mathcal{L}_{\text{PINC}} = \mathcal{L}_{\text{int}} + \mathcal{L}_{\text{diag}} + \mathcal{L}_{\text{grad}}$ gathers benefits of each component.

Overall both classes of methods greatly improve performance on physical losses when trained on $\mathcal{L}_{\text{PINC}}$, while slightly decreasing $\bm{f}$ PSNR.
The degradation in reconstruction observed for neural fields is connected to the interpretation of the physical loss scaling behaviors (\cref{sec:pinn_scaling}): as minimizing $\mathcal{L}_{\text{PINC}}$ shifts the modes to ones relevant for integrals and diagnostics, some of the dominant ones of $\bm{f}$ become less represented and the decoded quality slightly degrades.
While neural field training is generally consistent, for autoencoders severe instabilities emerge when training jointly on $\mathcal{L}_{\text{recon}}+\mathcal{L}_{\text{PINC}}$.
Our EVA finetuning procedure is consistently outperforming and more stable than directly training on $\mathcal{L}_{\text{PINC}}$ (bottom of \cref{tab:pinn_ablations}).

\subsection{Representation Space Experiments}
\label{sec:interp}
\begin{table}[h]
    \centering
    \small
    \renewcommand{\arraystretch}{1.2}
    \setlength{\tabcolsep}{1pt}

    \begin{minipage}[t]{0.54\linewidth}
        \centering
        \caption{Hybrid compression.\label{tab:hybrid}}
        \begin{tabular}{lcccc}
            \toprule
            Metric & & \texttt{ZFP} & \texttt{ZipNN} \\
            \midrule
            Extra CR & & 2.1$\times$ & 1.2$\times$ \\
            $\Delta$ PSNR ($\bm{f}$) & $\uparrow$ & -2e-4\% & 0\% \\
            $\Delta$ L1($Q$) & $\downarrow$ & +8e-3\% & 0\% \\
            $\Delta$ L1($\bm{{\phi}}$) & $\downarrow$ & +9.5\% & 0\% \\
            \bottomrule
        \end{tabular}
    \end{minipage}%
    \hfill
    \begin{minipage}[t]{0.44\linewidth}
        \centering
        \caption{Latent interpolation.\label{tab:interp}}
        \begin{tabular}{lc}
            \toprule
            Model & PSNR \\
            \midrule
            Extremes & 16.7 \\
            $\bm{f}$ (data) & 19.6 \\
            NF (weights) & 18.9 \\
            VQ-VAE (latents) & 20.5 \\
            \bottomrule
        \end{tabular}
    \end{minipage}
    \vspace{-8px}
\end{table}

\textbfp{Hybrid compression}
Neural methods can further improve the compression rate when coupled with traditional techniques on the learned representations.
Similarly to how data can be compressed into a low dimensional encoding, model parameters and autoencoder features are redundant and also lie on a lower-dimensional manifold. This is related to pruning \citep{lecun1990optimal, han2015deep}, network compression \citep{hershcovitch2024zipnn}, and the lottery ticket hypothesis \citep{frankle2019lotterytickethypothesisfinding}.

Improved compression can be achieved either with (lossless) entropy coding \citep{hershcovitch2024zipnn} or (lossy) quantization methods \citep{lindstrom2014fixed}. We apply both to neural fields and present findings in \cref{tab:hybrid}. 
ZipNN is lossless and does not induce any performance change, while providing a modest CR boost. 
ZFP is lossy with (tolerance of $10^{-3}$), leading to some performance degradation and $2.1\times$ CR.
Both results are averaged on 60 random samples from 10 trajectories.
We also show \texttt{NF + ZFP} in \cref{fig:scaling}. It closely follows the slope of \texttt{NF} but is right-shifted, achieving better CR.
Notably, at the higher regimes they appear to converge suggesting diminishing returns.
%
As an utmost example, we apply entropy coding on the VQ-VAE indices, raising the compression to 121492$\times$ (see \cref{app:entropyencoding}).

\textbfp{Latent (and weight space) interpolation}
Representational consistency and compactness over different snapshots is a desired property of compression methods.
It enables temporal coarsening \citep{ohana2024well, toshev2023lagrangebenchlagrangianfluidmechanics} by interpolation in weight/latent space resulting in additional gains in CR as not every single snapshot needs to be compressed.
To this end, we design an experiment to assess whether \ac{PINC} models exhibit representational consistency across time.
We encode two \textit{extremes} $\bm{f}_a, \bm{f}_b$ separated by $\Delta T$ and reconstruct intermediates $\bm{f}_t$ for $t = a, a+\mathrm{d} T, \dots, b$ by linearly interpolating the representations (latents or weights) $Z_{\bm{f}_a}$ and $Z_{\bm{f}_b}$.

For standard autoencoders, latent-space interpolation is a common practice \citep{berthelot2018understanding}. In the case of VQ-VAEs, the latents are interpolated before quantization to produce more accurate reconstructions.
It is not as straightforward for neural fields, as the parameters are not necessarily canonically ordered and exhibit various neuron symmetries \citep{HECHTNIELSEN1990129, NEURIPS2022_4df3510a}.
To address this, we use a shared \textit{meta neural field} trained on all extremes before finetuning it on each of them separately, improving weight alignment.
This is similar to the initialization strategy used by \citet{deluigi2023deeplearningimplicitneural} and \citet{erkoç2023hyperdiffusiongeneratingimplicitneural} to generate an aligned dataset of neural fields.

\cref{tab:interp} provides compelling evidence that linearly interpolating in representation space improves over simply taking the extremes, and approximates linear interpolation in data space. It shows time coarsening by evaluating on middle snapshot $t_m = t_l + \frac{\Delta T}{2}$. Interpolating representations is comparable to data-space and has more information than the extremes. Results are averaged on 50 (unseen) midpoints on 10 trajectories, with $\Delta T = 8R/V_\mathrm{r}$.
\cref{fig:interp_slice} illustrates intermediate reconstructions over time as progressive interpolation between $Z_{\bm{f}_a}$ and $Z_{\bm{f}_b}$.
However, because the underlying simulations are highly nonlinear accurate linear interpolation is unlikely, hence the low reported PSNR.
Regardless, we reckon that these results shows that learned representations are compact and self-consistent over time.

\section{Conclusions}

Our study provides compelling evidence that \acf{PINC} improves compression rate while maintaining underlying characteristics for gyrokinetic simulations of plasma turbulence.
This is achieved by constraining training to maintain integral quantities and spectral shapes across key dimensions of the 5D fields.
We anticipate that this approach can potentially be extended to other scientific domains, enabling practitioners to store compressed simulations that accurately capture specified physical phenomena across time and space, something previously infeasible due to storage requirements.
These tools could considerably improve data accessibility and transfer, accelerating research across scientific communities.

The compression methods presented in this work could be combined with \textit{neural operators}, nonlinearly evolving them in time. 
A major benefit of this is a significant reduction in dataset size required to train a surrogate model.
Orthogonally, exploring physics-informed "functasets" \citep{dupont2022functa,jo2025pdefunctaspectrallyawareneuralrepresentation} could be a valuable direction to further improve compression of neural fields for transient simulations and enable in-transit processing of data.
Related approaches in this regard include continual learning \citep{yan2021continual, woo2025metacontinuallearningneuralfields}, and in general ways to incorporate temporal dynamics into the training to enable on-the-fly (\textit{in-situ}) compression of simulation snapshots.

\textbfp{Limitations}
First, temporal information is not incorporated during \ac{PINC} training, which we expect to especially improve on compression ratio and temporal consistency.
Due to the computational complexity of training neural fields and especially autoencoders, this is left to future work.
Second, the computational requirements are substantial, mirrored in the training times (\cref{tab:timing}).
Even for neural fields, compression times are rather high and a consumer grade GPU is required.
Finally, the proposed physics-informed losses are specific to gyrokinetics, limiting transferability to other scientific areas beyond plasma physics.
Concurrent neural compression works, such as \citet{mo2022pivqae} for fluid dynamics and \citet{cranganore2025einsteinfieldsneuralperspective} for General Relativity, are also problem-specific.
To our knowledge there is no general loss reformulation that is applicable to any problem, and we leave generalizing \ac{PINC} to other domains as future work.
We postulate that the strategies and methodologies used for gyrokinetics-\ac{PINC}, for example the stabilization with EVA finetuning used for the autoencoders, can be successfully applied to other sources.

\section*{Impact Statement}
The potential impact of our work extends beyond machine learning, and the main purpose is to facilitate scientific discovery in computational plasma physics and general numerics. It does not warrant further discussion on the ethics.

\section*{Reproducibility statement}
Training and experiment code is submitted as a zip file in the supplementary materials. It contains autoencoders, neural fields and baseline implementation, as well as the configuration files used to obtain the paper results. The readme briefly outlines the code structure and describes how to start autoencoder/neural field training runs.
Some further information on training is already present in the Method and Results sections, as well as dedicated sections in the Appendix.
Unfortunately, the dataset used to train the autoencoders is not easily distributable due to its size. It was generated with the GKW \citep{PEETERS_GKW_2009} flux tube gyrokinetic numerical solver, as detailed in \cref{app:dataset}. A template for the configuration file used by GKW to start a run is included in the supplementary materials (\texttt{data\_generation/} directory).
Parameter ranges used to generate the dataset are included both in the supplementary as well as in \cref{app:dataset} for transparency.
We release on huggingface a validation dataset along with neural field weights and autoencoder checkpoints \href{https://huggingface.co/datasets/gerkone/pinc_gkw}{at this link}. These samples allow to reproduce the main results displayed in \cref{tab:results}.

\nocite{langley00}

\bibliography{references}
\bibliographystyle{icml2026}

\newpage
\appendix
\onecolumn

\section{Gyrokinetics}
\label{app:gyrokinetics}
Gyrokinetics \citep{frieman1982nonlinear, Krommes_gyrokinetics_2012, PEETERS_GKW_2009} is a reduced form of Plasma kinetics that is computationally more efficient and can be use to locally simulate Plasma behavior within a so-called \textit{flux tube} in the torus.
Local gyrokinetics is a theoretical framework to study plasma behavior on perpendicular spatial scales comparable to the gyroradius, i.e., the radius of  circular motion exhibited by charged particles in a magnetic field, and frequencies much lower than the particle cyclotron frequencies, i.e., the  frequency at which charged particles spiral around magnetic field lines due to the Lorentz force.
Gyrokinetics models the time evolution of electrons and ions
via the distribution function $\bm{f}$, which is based on 3D coordinates, their parallel and perpendicular velocities, together with the angle w.r.t. the field lines. However, the latter dimension is averaged out by modelling only the so-called guiding center of a particle instead of its gyral movement.
Furthermore, instead of modelling the perpendicular velocity, usually only its magnitude is considered, which is also referred to as the magnetic moment $\mu$.
Hence, the 5D gyrokinetic distribution function can be written as $\bm{f} = \bm{f}(k_{x},k_{y},s, v_{\parallel}, \mu)$, where
$k_x$ and $k_y$ are spectral coordinates,
$s$ is the toroidal coordinate along the field line, and $v_{\parallel}$ the parallel velocity component.
The perturbed time-evolution of $\bm{f}$, for each species (ions and electrons), is governed by 
\begin{equation}
\underbrace{\colorbox{blue!10}{$\displaystyle
\frac{\partial \bm{f}}{\partial t} + (v_\parallel {\Bb} + {\Bv}_D) \cdot \nabla \bm{f}   
-\frac{\mu B}{m} \frac{\BB \cdot \nabla B}{B^2} \frac{\partial \bm{f}}{\partial v_\parallel}$}}_\text{Linear}
\; + \;
\underbrace{\colorbox{red!30}{$\displaystyle
\smash{{\Bv}_\chi \cdot \nabla \bm{f}}
\vphantom{\frac{\partial \bm{f}}{\partial v_\parallel}}$}}_\text{Nonlinear} 
 = S \ , 
\label{gyrovlas}
\end{equation}
where $v_\parallel \Bb$ is the motion along magnetic field lines, $\Bb = \BB / B$ is the unit vector along the magnetic field $\BB$ with magnitude $B$\footnote{We adopt uppercase notation for vector fields $\BE$ and $\BB$ to adhere with literature.}, $\Bv_D$ the magnetic drift due to gradients and curvature in $\BB$, and $\Bv_\chi$ describes
drifts arising from the $\BE \times \BB$ force, a key driver of plasma dynamics.
Finally, $S$ is the source term, the external supply of energy.
The term $\Bv_\chi \cdot \nabla \bm{f}$ models the nonlinear interaction between the distribution function $\bm{f}$ and its velocity space integral $\bm{\phi}$, and it describes turbulent advection. The resulting nonlinear coupling constitutes the computationally most expensive term. 

\subsection{Derivation of the Gyrokinetic equation}
\label{app:gyrokinetics_derivation}

We begin with the Vlasov equation for the distribution function $\bm{f}(\Br, \Bv, t)$:
\begin{equation}
\frac{\partial \bm{f}}{\partial t} + \mathbf{v} \cdot \nabla \bm{f} + \frac{q}{m} \left( \BE + \Bv \times \BB \right) \cdot \nabla_v \bm{f} = 0
\end{equation}

The Vlasov equation describes the conservation of particles in phase space in the absence of collisions.
Here, $\Br=(x,y,z)$ and $\Bv=(v_x, v_y, v_z)$ correspond to coordinates in the spatial and the velocity domain, respectively.
Hence the Vlasov equation is a 7D (including time) PDE  representing the  density of particles in phase space at position $\Br$, velocity $\Bv$, and time.
The term $\nabla_{\Bv} \bm{f}$ describes the response of the distribution function to accelerations of particles and $\frac{q}{m} \left( \mathbf{E} + \mathbf{v} \times \mathbf{B} \right)$ denotes the Lorentz force, which depends on particle charge $q$ and mass $m$, as well as electric field $\BE$ and magnetic field $\BB$.
Finally, the advection (or convection) term $\Bv \nabla \bm{f}$ describes transport of the distribution functon through space due to velocities.

To derive the \textit{gyrokinetic equation}, we transform from particle to guiding center coordinates $(\mathbf{R}, v_\parallel, \mu, \theta)$, where $\mu = \frac{m v_\perp^2}{2B}$ is the magnetic moment, $\theta$ the gyrophase (particle position around its guiding center as it gyrates along a field line), and $\BR$ is the coordinate of the guiding center.

Assuming the time scale $L$ at which the background field changes is much longer than the gyroperiod (small Larmor radius $\rho \ll L$), we can \textit{gyroaverage} to remove the dependency on the gyrophase $\theta$:
\begin{equation}
\frac{\partial \bm{f}}{\partial t} + \dot{\BR} \cdot \nabla \bm{f} + \dot{v}_\parallel \frac{\partial \bm{f}}{\partial v_\parallel} = 0
\end{equation}

\subsubsection{Linear Terms}

The unperturbed (background) motion of the guiding center is governed by:
\begin{align}
\dot{\BR} &= v_\parallel \Bb + \Bv_D \\
\dot{v}_\parallel &= -\frac{\mu}{m} \Bb \cdot \nabla \BB
\end{align}

Here,  $\mathbf{b} = \BB/B$ is the unit vector along the magnetic field, and $\Bv_D$ represents magnetic drifts. Substituting into the kinetic equation yields
\begin{equation}
\frac{\partial \bm{f}}{\partial t} + (v_\parallel \Bb + \Bv_D) \cdot \nabla \bm{f} 
- \frac{\mu}{m} \Bb \cdot \nabla \BB \frac{\partial \bm{f}}{\partial v_\parallel} = 0
\end{equation}

We can express the magnetic gradient term using:
\begin{equation}
    \Bb \cdot \nabla \BB = \frac{\BB \cdot \nabla \BB}{B}    
\end{equation}
so that:
\begin{equation}
\frac{\mu}{m} \Bb \cdot \nabla \BB = \frac{\mu B}{m} \frac{\BB \cdot \nabla B}{\BB^2}
\end{equation}

\subsubsection{Nonlinear Term}

Fluctuating electromagnetic potentials $\delta \bm{\phi}, \delta \BA$ induce E$\times$B and magnetic flutter drifts. 
We define the gyroaveraged generalized potential as
\begin{equation}
    \chi = \langle \bm{\phi} - \frac{v_\parallel}{c} A_\parallel \rangle,    
\end{equation}
where $\BA_\parallel$ is the parallel component of the vector potential, $\langle\cdot\rangle$ denotes the gyroaverage, and $c$ is the speed of light, which is added to ensure correct units. $\bm{\phi}$ is the electrostatic potential, the computation of which involves an integral of $\bm{f}$ over the velocity space (see eq. 1.41 in the GKW manual \footnote{\url{https://bitbucket.org/gkw/gkw/src/develop/doc/manual/}} for a complete description). 

This gives rise to the drift
\begin{equation}
    \mathbf{v}_\chi = \frac{c}{B} \mathbf{b} \times \nabla \chi, 
\end{equation}
and yields the nonlinear advection term $\mathbf{v}_\chi \cdot \nabla \bm{f}$ .

\subsubsection{Final Equation}

We arrive at the gyrokinetic equation in split form:
\begin{equation}
\frac{\partial \bm{f}}{\partial t} + (v_\parallel \mathbf{b} + \mathbf{v}_D) \cdot \nabla \bm{f}   
-\frac{\mu B}{m} \frac{\mathbf{B} \cdot \nabla B}{B^2} \frac{\partial \bm{f}}{\partial v_\parallel}
\; + \;
\mathbf{v}_\chi \cdot \nabla \bm{f}
\; = \;
S
\end{equation}

Here, \( S \) represents external sources, collisions, or other drive terms.
To enhance the tractability of \cref{gyrovlas}, the distribution function $\bm{f}$ is usually split into equilibrium and perturbation terms
\begin{equation}
    \bm{f} = \bm{f}_0 + \delta \bm{f} = \underbrace{\colorbox{blue!10}{$\displaystyle \bm{f}_0 - \frac{Z\bm{\phi}}{T}\bm{f}_0$}}_\text{Adiabatic}\; + \; \underbrace{\colorbox{red!30}{$\displaystyle \frac{\partial h}{\partial t}$}}_\text{Kinetic} \ ,
\end{equation}
where $\bm{f}_0$ is a background or equilibrium distribution function, $T$ the particle temperature, $Z$ the particle charge, $\bm{\phi}$ the electrostatic potential, and $\delta f$ the total perturbation to the distribution function, which comprises of \textit{adiabatic} and \textit{kinetic} response. 
The adiabatic term describes rapid and passive responses to the electrostatic potential that do not contribute to turbulent transport, while the kinetic term governs irreversible dynamics that facilitate turbulence. Numerical codes, such as GKW \citep{PEETERS_GKW_2009}, rely on solving for $\delta f$ instead of $\bm{f}$.
A common simplification is to assume that electrons are adiabatic, which allows us to neglect the kinetic term in the respective $\delta \bm{f}$.
Hence, the respective $\bm{f}$ for electrons ($\bm{f}_e$) does not need to be modelled, effectively halving the computational cost.

\section{Dataset}
\label{app:dataset}
The simulations used for both the autoencoder training (26 trajectories) and the evaluation (10 trajectories) are generated with the numerical code \emph{GKW} \citep{PEETERS_GKW_2009}.
They are sampled by varying four parameters: $R/L_t$, $R/L_n$, $\hat{s}$, and $q$, which significantly affect emerging turbulence in the Plasma.
\begin{itemize}
    \item $R/L_t$ is the ion temperature gradient, which is the main driver of turbulence.
    \item $R/L_n$ is the density gradient, whose effect is less pronounced. It can have a stabilizing effect, but can sometimes also lead to increased turbulence.
    \item $\hat{s}$ denotes magnetic shearing, hence it usually has a stabilizing effect as more magnetic shearing results in better isolation of the Plasma.
    \item $q$ denotes the so-called safety factor, which is the inverse of the rotational transform and describes how often a particle takes a poloidal turn before taking a toroidal turn.
\end{itemize}
We specify the ranges for sampling the four parameters as $R/L_T \in [1,12]$, $R/L_n \in [1,7]$, $q \in [1,9]$, and $\hat{s} \in [0.5,5]$.
Additionally, we also vary the noise amplitude of the initial condition (within $[1e-5, 1e-3]$).

To make storage more feasible, simulations are time-coarsened by saving snapshot every 60. Each GKW run with the specified configurations takes around $\sim$6 hours (76 cores, Intel Ice Lake 4.1GHz CPU) and $\sim$60GBs of storage.

\section{Implementation details}
\label{app:implementation}

\subsection{Metrics}
\label{app:metrics}
We evaluate reconstruction with spatial and physical metrics. Since gyrokinetic data is complex-valued, we can also apply complex-generalizations of common metrics.

\textbfp{Complex L1 Loss}
Given two complex-valued fields \( z_1, z_2 \in \mathbb{C}^N \), the complex L1 loss is:
\[
\mathrm{cL1}(z_1, z_2) = \left\langle |\Re(z_1 - z_2)| + |\Im(z_1 - z_2)| \right\rangle = \left\langle \left| z_1 - z_2 \right|_1 \right\rangle
\]
where \( \langle \cdot \rangle \) denotes the average over all dimensions and \( |\cdot|_1 \) is the L1 norm of the complex difference.

\textbfp{Wasserstein Distance}
The Wasserstein distance measures the minimum cost of transforming one probability distribution into another, where the cost is proportional to the distance the probability mass must be moved. It provides a meaningful metric to compare distributions even when they have non-overlapping support, making it particular useful in machine learning and optimal transport problems. In our case, we normalize the spectra so that their total sum is one, ensuring they represent comparable probability distributions.
The Wasserstein distance between two probability distributions $P$ and $Q$ is defined as:
\[
W_p(P, Q) = \left( \inf_{\gamma \in \Gamma(P, Q)} \int \|x - y\|^p \, d\gamma(x, y) \right)^{\frac{1}{p}}
\]

\textbfp{Peak Signal-to-Noise Ratio}
Peak signal-to-noise ratio (PSNR) quantifies the ratio between the maximum possible power of a signal and the power of noise corrupting its representation, typically expressed in decibels (dB) due to the wide dynamic range of signals.
\[
\mathrm{PSNR}(x_1, x_2) = 10 \cdot \log_{10} \left( \frac{ \max(x_1)^2 }{ \mathrm{MSE}(x_1, x_2) } \right)
\]

where $\mathrm{MSE}(x_1, x_2)$ is the mean squared error between the real-valued fields $x_1$ and $x_2$.
The PSNR for complex-valued fields we defined as:
\[
\mathrm{cPSNR}(z_1, z_2) = 10 \cdot \log_{10} \left( \frac{ \max(|z_1|)^2 }{ \mathrm{cMSE}(z_1, z_2) } \right)
\]

\textbfp{Bits Per Pixel (BPP)}
The BPP measures compression efficiency. 
Given a discrete representation of a field $z$ and its compressed encoding, the bits per pixel is defined as
\[
\mathrm{BPP} = \frac{\text{Total number of bits used to encode } z}{\text{Number of spatial points in } z}.
\]

\textbfp{Optical Flow and End-Point Error (EPE)}
\label{app:temporal_metrics}
Optical flow estimates the apparent motion between consecutive frames by computing spatial gradients and temporal derivatives.
Applying the standard video algorithms to 5D data by reshaping or flattening dimensions introduces physical discontinuities and artificial flow boundaries, resulting in non-existent "ghost" flows.
Instead, we extend the popular Horn–Schunck method \citep{HORN1981185} to 5D, which computes the optical flow field directly on the 5D grid. In this case, the optical flow is a velocity vector field $\mathbf{u} = (u_{v_{\parallel}}, u_{\mu}, u_s, u_x, u_y)$, resulting in 5 components per grid location.

For two consecutive frames $x(t)$ and $x(t+1)$, we compute the temporal derivative $\delta_t x$ and the spatial gradient vector $\nabla x = (\partial_{v_{\parallel}} x, \partial_{\mu} x, \partial_{s} x, \partial_{x} x, \partial_{y} x)$ using finite differences on the temporal mid-point average.
The full Horn-Schunck method enforces global smoothness by iteratively minimizing an energy functional, balancing brightness constancy with a smoothness term weighted by $\alpha$. This formulation yields the following iterative update rule for the $k$-th iteration:
\begin{equation*}
\mathbf{u}^{(k+1)} = \bar{\mathbf{u}}^{(k)} - \frac{\nabla x \left( \nabla x \cdot \bar{\mathbf{u}}^{(k)} + \Delta x \right)}{\alpha^2 + \|\nabla x\|^2},
\end{equation*}
where $\bar{\mathbf{u}}^{(k)}$ represents the local average of the flow vectors from the previous iteration. This average is computed via 5D convolution using a star kernel $K$. It assigns a uniform weight of $1/10$ to the $10$ immediate neighbors ($2$ neighbors, $5$ dimensions) along the spatial dimensions. The pseudocode we use for the 5D Horn-Schunck is in Algorithm \ref{lst:hs_5d}.
Without the smoothing step it's the special case of \textit{normal flow}, capturing motion strictly in the direction of the gradient (e.g. disregarding diagonal shifts).

\begin{lstlisting}[caption={Generalized 5D Horn--Schunck}, label={lst:hs_5d}]
def horn_schunck($x$ (5+1D data), $\alpha$ (regularization), $N$ (iters)):
    $\delta_t x$ = $x_{t+1} - x_t$  # temporal gradient (assume regular frame-rate)
    $\nabla x$ = gradient($0.5(x_t + x_{t+1})$)  # spatial gradient
    $D = \alpha^2 + \|\nabla x\|^2$  # normalization factor
    # iterative smoothing
    kernel = star(neighbors=2 $\cdot$ 5, kernel_size=3)
    $u \leftarrow 0$
    for $k = 1$ to $N$:
        # global smoothness via local averaging
        $\bar{u}$ = convolve($u$, kernel)
        # Update flow (project temporal change onto spatial gradient)
        project = $(\nabla x \cdot \bar{u} + x_t) / D$
        $u \leftarrow \bar{u} - \nabla x \cdot \text{project}$
    return $u$
\end{lstlisting}

Given two sequences of $N$ frames, the End-Point Error (EPE) \citep{Baker2011epe} is defined as the mean squared Euclidean difference between the predicted flow vectors $\mathbf{F}_{pred}^{(i,j)}$ and the target flow vectors $\mathbf{F}_{tgt}^{(i,j)}$ over all $M$ grid points and time steps:
\begin{equation*}
\mathrm{EPE} = \frac{1}{N \cdot M} \sum_{i=1}^N \sum_{j=1}^M \|\mathbf{F}_{pred}^{(i,j)} - \mathbf{F}_{tgt}^{(i,j)} \|_2^2.
\end{equation*}

\subsection{Traditional compression}
\label{app:trad}
In the following paragraphs we briefly describe how the traditional compressions were implemented.

\textbfp{ZFP Compression}
ZFP \citet{lindstrom2014fixed} is a compression library for numerical arrays designed for fast random access. It partitions the data into small blocks (typically $4\times 4 \times 4$ elements for 3D data) and transforms them into a decorrelated representation using an orthogonal block transform. The transformed coefficients are quantized according to a user-specified tolerance, then entropy-coded to produce a compact bitstream. High-speed random access and both lossy and lossless are possible, making ZFP a very common choice for scientific data storage.

We rearrange $\bm{f}$ into a 3D array as $((v_{\parallel}\times \mu) \times (s\times y) \times x)$ for ZFP block-based compression scheme (up to 3D), and compress with ZFP with a specified absolute error tolerance. The compressed representation is a compact byte representation. Reconstruction is performed by decompressing with ZFP and reshaping the output back to the original tensor layout.

\textbfp{Wavelet Compression}
Discrete wavelet transform (DWT) is applied using the level 1 Haar wavelet. The multi-dimensional array is decomposed into wavelets (coefficient and slices). To achieve lossy compression, coefficients are pruned based on a fixed threshold dependent on the desired compression ratio, effectively discarding small high-frequency components. Reconstruction is performed by inverting the DWT.

\textbfp{Principal Component Analysis Compression}
$\bm{f}$ is reshaped into a 2D array $((v_{\parallel} \cdot \mu \cdot s) \times (x \cdot y))$, by rearranging together the velocity space $v_{\parallel},\mu$ with the field line $s$ and the spatial coordinates $x,y$. PCA is applied on the flattened spatial components, retaining a fixed number of principal components dependent on the desired compression ratio ($N=2$ for $1,000\times$ from \cref{tab:results}). The compressed representation consists of the principal components, the mean vector, and the explained variance. Reconstruction is achieved by projecting back to the original space, followed by reshaping to the original dimensions.

\textbfp{JPEG2000 Compression}
$\bm{f}$ is first reshaped into a 2D image-like representation of shape $((v_{\parallel}\cdot \mu \cdot s) \times (x \cdot y))$, by flattening the velocity space and spatial dimensions. Each channel is independently normalized to the $[0,1]$ range and quantized to 16-bit unsigned integers. The images are then encoded using the JPEG2000 standard \citep{jpeg2000} at a target quality factor $Q$ that determines the compression ratio. The compressed representation consists of the codestream size and channelwise normalization statistics (minimum and maximum). Reconstruction is performed by decoding the JPEG2000 bitstream, rescaling back to floating-point values, and unflattening back to the original tensor dimensions.

\subsection{VAPOR}
\label{app:vapor}
VAPOR \citep{choi2021neural} combines a VQ-VAE \citet{oord2017neural} compressor and a a Fourier Neural Operator (FNO) \citep{li2021fourierneuraloperatorparametric} Refiner sequentially. The VQ-VAE provides extreme compression by reducing the size of the original data, and the FNO Refiner then refines the VQ-VAE's coarse output to restore fidelity, achieving both high compression and high accuracy.
We utilize a VQ-VAE with Exponential Moving Average (EMA) updates to compress the data $\bm{f}$. This forms the first stage of the overall architecture.
The FNO refiner stage uses a residual structure to efficiently learn and apply the high-frequency corrections needed to match the ground-truth solution, taking the VQ-VAE initial reconstruction as input.

Finally, a core component of \citet{choi2021neural} is the specialized physics loss $\mathcal{L}_{\text{physics}}$, employed to enforce conservation laws.
This loss computes the MSE between the predicted and ground-truth values of density, momentum, and energy:
\begin{align*}
\mathcal{L}_{\text{physics}} \:=\: &
\operatorname{MSE}\!\left(
\sum_{v_\parallel,\,v_\perp} \bm{f}_{\mathrm{pred}},\;
\sum_{v_\parallel,\,v_\perp} \bm{f}_{\mathrm{gt}}
\right)
\:+\:
\operatorname{MSE}\!\left(
\sum_{v_\parallel,\,v_\perp} \bm{f}_{\mathrm{pred}}\, v_\parallel,\;
\sum_{v_\parallel,\,v_\perp} \bm{f}_{\mathrm{gt}}\, v_\parallel
\right)
\\
+&
\operatorname{MSE}\!\left(
\sum_{v_\parallel,\,v_\perp} 
\bm{f}_{\mathrm{pred}}\, \tfrac12 m_s v_\parallel^2,\;
\sum_{v_\parallel,\,v_\perp} 
\bm{f}_{\mathrm{gt}}\, \tfrac12 m_s v_\parallel^2
\right).
\end{align*}
This loss is added to the standard reconstruction and VQ losses during training to obtain the final VAPOR loss: $\mathcal{L} = \mathcal{L}_{\text{recon}} + \mathcal{L}_{\text{VQ}} + \mathcal{L}_{\text{physics}}$. Results in \cref{tab:vapor_results}.

\begin{table}[b]
\renewcommand{\arraystretch}{1.2}
\scriptsize
\caption{VAPOR results on the same setup as \cref{tab:results}. \label{tab:vapor_results}}
\centering
\resizebox{\textwidth}{!}{%
\begin{tabular}{lr|cc|cc|cc}
\toprule
         &       & \multicolumn{2}{c}{Compression $\bm{f}$}                             & \multicolumn{2}{c}{Integrals $Q,\bm{\phi}$}                                              & \multicolumn{2}{c}{Turbulence $Q^{\text{spec}}, k_y^{\text{spec}}$}                                        \\ \midrule
         & CR 
         & PSNR $\uparrow$  & $\text{EPE}$ $\downarrow$  &
         $\text{L1}(Q)$ $\downarrow$ & $\text{PSNR}(\bm{\phi})$ $\uparrow$ &
         $\text{WD}(\overline{k_y^{\text{spec}}})$ $\downarrow$ & $\text{WD}(\overline{Q^{\text{spec}}})$ $\downarrow$ \\\midrule
VAPOR & 64$\times$ & 29.52$_{\pm 1.36}$ & 0.123$_{\pm 0.07}$ & 90.75$_{\pm 48.41}$ & -13.96$_{\pm 5.97}$ & 0.020$_{\pm 0.01}$ & \underline{0.010$_{\pm 0.00}$} \\\midrule
PINC-NF  & 1167$\times$ & \underline{35.76$_{\pm 1.38}$} & \underline{0.037$_{\pm 0.02}$} & \textbf{2.18$_{\pm 8.33}$} & \textbf{13.50$_{\pm 4.44}$} & \textbf{0.006$_{\pm 0.00}$} & 0.015$_{\pm 0.00}$ \\
AE + EVA & 716$\times$ & 35.64$_{\pm 2.03}$ & 0.063$_{\pm 0.05}$ & \underline{15.01$_{\pm 16.42}$} & 6.72$_{\pm 4.98}$ & 0.016$_{\pm 0.01}$ & 0.012$_{\pm 0.01}$ \\
\bottomrule
\end{tabular}
}
\end{table}

\subsection{Autoencoders}
\label{app:swin}

\begin{wrapfigure}{r}{0.42\textwidth}
    \centering
    \includegraphics[width=\linewidth]{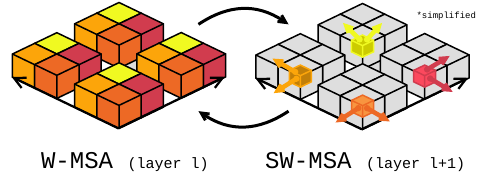}
    \vspace{-12px}
    \caption{5D swin attention.}
    \label{fig:swin}
    \vspace{-8px}
\end{wrapfigure}

\textbfp{Architecture and Conditioning}
The autoencoder and VQ-VAE baselines are built on a 5D Swin Transformer architecture \citep{galletti20255dneuralsurrogatesnonlinear, paischer2025gyroswin}, which extends the shifted window attention mechanism to handle high-dimensional scientific data. \cref{fig:swin} illustrates the 5D windowed multi-head self-attention (W-MSA) and shifted windowed multi-head self-attention (SW-MSA) layers, where blocks of the same color indicate the receptive field of local attention within each window. Our implementation incorporates several stability and performance enhancements: gated attention mechanisms \citep{qiu2025gatedattentionlargelanguage} for improved training stability, combined positional encodings using both Relative Positional Bias \citep{liu_swin_2021} and Rotary Position Embedding (RoPE) \citep{su2023roformerenhancedtransformerrotary} to capture spatial relationships across all five dimensions, and GELU activations \citep{hendrycks2023gaussianerrorlinearunits} throughout the network.
Each model uses four Swin blocks with 16 attention heads, followed by a single downsampling level before the bottleneck. All models are conditioned on four key gyrokinetic parameters: the ion temperature gradient ($ R/L_t$), density gradient ($R/L_n$), magnetic shear ($\hat{s}$), and safety factor ($q$). Conditioning is implemented via DiT-style modulation \citep{peebles2023dit}, where conditioning embeddings provide scale, shift, and gating parameters for each transformer layer, enabling physics-aware feature adaptation.

\textbfp{Data Preprocessing}
The 5D distribution function $[v_\parallel, \mu, s, x, y]$ is represented as complex values with real and imaginary components, initially providing two channels. We apply two key preprocessing steps that affect the channel structure. First, we decomposes each field into zonal flow ($k_y=0$ mode) and turbulent fluctuation components by computing the mean across the $k_y$ dimension and concatenating the zonal flow and turbulent fluctuation, doubling the channels to four. This separation is essential as zonal flows exhibit fundamentally different physics from turbulent modes.
Second, we reshape the magnetic moment dimension $\mu$, into the channel dimension, expanding from four to 32 channels. This allows independent processing of each $\mu$ slice.

\textbfp{Compression Configurations}
We evaluate multiple compression ratios by varying patch and window sizes. For autoencoders, three configurations achieve compression ratios of 302, 1208, and 2865 using patch sizes $(2,0,2,5,2)$, $(4,0,2,5,4)$, and $(6,0,3,5,6)$ with corresponding window sizes $(8,0,4,9,8)$, $(4,0,4,9,4)$, and $(6,0,6,9,6)$. The zero in the second position corresponds to the $\mu$ dimension, which is not spatially patched due to the decoupling preprocessing step. All variants use latent dimension 1024, and compress in a last linear projection to the bottleneck dimension of 32.

\textbfp{VQ-VAE Variants}
VQ-VAE uses the same spatial compression configurations but replaces the continuous bottleneck with vector quantization using the implementation from \texttt{vector-quantize-pytorch}\footnote{\url{https://github.com/lucidrains/vector-quantize-pytorch}}. The bottleneck projects to 128-dimensional embeddings, which are quantized using a codebook of 8192 vectors (see \cref{tab:vqvae_params} for complete hyperparameters). The codebook uses exponential moving average updates with a decay rate of 0.99 and employs entropy regularization to encourage codebook utilization. This yields much higher compression ratios of 19342, 25789, and 77368 for the three spatial configurations, as quantized codes can be stored as integers (int16 for codebook size of 8192) rather than float32 values.

\begin{wraptable}{r}{0.45\textwidth}
    \centering
    \caption{VQ-VAE vector quantization hyperparameters.}
    \label{tab:vqvae_params}
    \begin{tabular}{ll}
        \toprule
        \textbf{Parameter} & \textbf{Value} \\
        \midrule
        Codebook size      & 8192      \\
        Embedding dimension & 128       \\
        Commitment weight  & 0.3       \\
        Codebook type      & Euclidean \\
        EMA decay          & 0.99      \\
        Entropy loss weight & 0.01      \\
        Dead code threshold & 2         \\
        \bottomrule
    \end{tabular}
\end{wraptable}

\textbfp{Training Strategy}
Training follows a two-stage approach to ensure stability. For all experiments we use Muon optimizer \citep{jordan2024muon} with a cosine scheduler and a minimum learning rate of $4 \times 10^{-6}$, and weight decay of $1 \times 10^{-5}$.  \textit{Stage 1} (200 epochs, batchsize=16, lr=$2 \times 10^{-4}$) trains the base autoencoder using only $\mathcal{L}_{\text{recon}}$ (cMSE).
\textit{Stage 2.1} (100 epochs, batchsize=16, lr=$2 \times 10^{-4}$) applies Explained Variance Adaptation (EVA) \citep{paischer2025eva}, which injects LoRA \citep{hu2022lowrank} weights ($r=64$, $\alpha=1$, $\rho=2.0$, $\tau=0.99$) into MLP layers while freezing the Stage 1 trained backbone. The loss function switches to cL1 for reconstruction ($\mathcal{L}_{\text{recon}}$ weighted by 10.0) and introduces physics-informed losses: integral losses ($\mathcal{L}_Q$, $\mathcal{L}_{\bm{\phi}}$) using scale normalization (scale is calculated over training dataset statistics), while spectral losses ($\mathcal{L}_{k_y}$, $\mathcal{L}_{Q^{\text{spec}}}$) employ sum-normalization followed by log-space L1 loss. All physics-informed loss terms are weighted equally at 1.0, with the VQ-VAE commitment loss also weighted by a factor of 10.0 to match the reconstruction weight. Critically, monotonicity constraints ($\mathcal{L}_{\text{iso}}$) are disabled. \textit{Stage 2.2} (20 epochs, batchsize=16, lr=$2 \times 10^{-4}$) continues with identical settings but enables monotonicity losses ($\mathcal{L}_{\text{iso}}(k_y^{\text{spec,pred}})$, $\mathcal{L}_{\text{iso}}(Q^{\text{spec,pred}})$) to enforce physical constraints only after stable physics-informed reconstruction is achieved.

\textbfp{Training Stabilization}
End-to-end training of autoencoders with physics-informed neural compression (\ac{PINC}) losses proves highly unstable due to the conflicting optimization objectives and varying loss magnitudes. The physics-informed terms ($\mathcal{L}_Q$, $\mathcal{L}_{\bm{\phi}}$, $\mathcal{L}_{k_y}$, $\mathcal{L}_{Q^{\text{spec}}}$) exhibit severe fluctuations during early training when reconstruction quality is poor, causing certain loss components to dominate the overall objective and destabilizing the learning process. This necessitates the staged training approach, where reconstruction capability is first established before introducing physics constraints.

\textbfp{Multi-objective Optimization Challenges}
We investigated several multi-objective optimization strategies to enable stable end-to-end training. Gradient normalization methods \citep{gradnorm2018}, while theoretically appealing, proved computationally prohibitive for our large-scale models, consistently causing out-of-memory errors during backpropagation. Conflict-Free Inverse Gradients (ConFIG) \citet{Liu2024ConFIG} attempts to resolve conflicting optimization objectives by computing gradient directions that minimize conflicts between tasks through least-squares solutions. However, ConFIG relies on computing stable gradient statistics over multiple training steps to determine optimal gradient directions. When physics-informed losses are computed on poorly reconstructed distribution functions, these losses exhibit extreme fluctuations that prevent ConFIG from establishing stable gradient statistics. The method's gradient balancing becomes ineffective when the underlying loss landscape is highly unstable, as the computed conflict-free directions become unreliable due to the volatile nature of the physics-informed terms during early training phases.

\textbfp{Hyperparameter Search Limitations}
The computational cost of autoencoder training further complicates optimization. Each full training run requires multiple days on high-end GPUs, making systematic hyperparameter search for end-to-end training impractical. The search space includes not only standard hyperparameters (learning rates, batch sizes, architectural choices) but also the relative weighting of several distinct loss components, creating a prohibitively large optimization landscape. This computational constraint reinforces the necessity of our staged approach, which reduces the hyperparameter search to manageable subspaces for each training phase.

\textbfp{Codebook usage and Entropy Encoding}
\label{app:entropyencoding}
The VQ-VAE quantizes the continuous latent space into discrete integer indices (\textit{codes}) ranging from $0$ to the codebook size. Standard storage uses fixed-width encoding $\log_2(8192) = 13$ bits per code.
However, empirical analysis reveals non-uniform usage: frequent codes dominate (common turbulent structures), while rare codes occur sporadically. This imbalance enables lossless compression via variable-length entropy coding. Our VQ-VAE achieves 71.4\% codebook utilization (5846/8192 entries). The sorted codebook frequencies follow Zipf's law, which suggest common flow patterns use frequent codes, while rare events retain dedicated codes.
Further, we measure this redundancy using Shannon entropy $-\sum_i p_i \log_2(p_i)$ where $p_i$ is the empirical probability of code $i$. Our dataset yields $H \approx 10.5$ bits, indicating that optimal encoding requires only 10.5 bits per code on average, compared to the 13-bit fixed-width baseline.
On our test set, Huffman encoding \citep{huffman1952encoding} achieves $1.56 \times$ additional compression over fixed-width storage, reducing average code length from 13 to 10.7 bits per code. Combined with VQ-VAE quantazation ($77368 \times$), the total pipelines achieves $121492 \times$ compression, going from 723.5GB (uncompressed) to 5.96MB (VQ-VAE + Huffman), instead of 9.32MB (VQ-VAE).

\subsection{Neural fields}
\label{app:nf}
Neural fields are trained by representing the distribution function as a continuous signal, taking coordinates as inputs. A dataset consists, for a given simulation, of the 5D density function $\bm{f}$ at a specific timestep, and the 5D grid coordinates of each cell. Data normalization is applied both to the field values and to the coordinates.

An MLP with SiLU activations \citep{elfwing2017sigmoidweightedlinearunitsneural}, 64 hidden dimension, five layers with skip connections and using a discrete hash to map matrix indices to learnable embeddings is optimized using AdamW \citep{loshchilov2019decoupledweightdecayregularization}, with cosine annealing learning rate scheduling decaying the learning rate from $5e-3$ to $1e-12$ and . Auxiliary optimizers can be used for additional integral losses, also with their scheduler that decays learning rate from $1e-5$ to $1e-12$. The neural field training loop iterates over batches of (2048) coordinates and field values. On a first pass of 20 epochs, the loss $\mathcal{L}_{recon}$ from \cref{eq:recon_loss} is fitted. Auxiliary integral losses are trained of such a pretrained model for 100 more epochs, with the whole 5D field as batch. 

\textbfp{ConFIG ablations}
We ablate multi-objective balancing methods such as Conflict-Free Inverse Gradients by \citet{Liu2024ConFIG} to attempt to stabilize training on the \ac{PINC} loss terms. \cref{tab:config_results} compares AdamW training (as reported in \cref{tab:results}) and neural fields complemented with momentum ConFIG with ordered loss selector. Results are similar, with regular AdamW achieving better physical losses and ConFIG being more stable overall.

\begin{table}[h]
\renewcommand{\arraystretch}{1.1}
\scriptsize
\caption{Ablations of NF trained with AdamW and Conflict-Free Inverse Gradients.
\label{tab:config_results}}
\centering
\resizebox{\textwidth}{!}{%
\begin{tabular}{lc|ccc|cc|cc}
\toprule
 &       & \multicolumn{3}{c}{Compression $\bm{f}$}                             & \multicolumn{2}{c}{Integrals $Q,\bm{\phi}$}                                              & \multicolumn{2}{c}{Turbulence $Q^{\text{spec}}, k_y^{\text{spec}}$}                                        \\ \midrule
 & CR    & L1  $\downarrow$  & PSNR $\uparrow$    & BBP $\downarrow$   & $\text{L1}(Q)$ $\downarrow$ & $\text{PSNR}(\bm{\phi})$ $\downarrow$ & $\text{WD}(\overline{k_y^{\text{spec}}})$ $\downarrow$ & $\text{WD}(\overline{Q^{\text{spec}}})$ $\downarrow$ \\ \midrule
\ac{PINC}-NF (AdamW)      & 1163$\times$ & 0.32 & 36.29 & 0.165 & \textbf{9.75} & \textbf{14.53} & \textbf{0.0057} & 0.0170 \\
\ac{PINC}-NF (SGD+ConFIG) & 1163$\times$ & \textbf{0.29} & \textbf{37.18} & 0.165 & 44.23 & 6.35 & 0.0164 & \textbf{0.0163} \\
\bottomrule
\end{tabular}
}
\end{table}

\textbfp{Neural field ablations}
\label{app:nf_ablations}
A broad range of architectures was explored, starting from SIREN \citep{sitzmann2020siren}, WIRE \citet{saragadam2023wire} and an MLP with different activations \citep{fukushima1969relu, hendrycks2023gaussianerrorlinearunits, elfwing2017sigmoidweightedlinearunitsneural}. \cref{tab:nf_ablations} summarizes the search space. 

\begin{table}[h]
\centering
\renewcommand{\arraystretch}{1.1}
\caption{Neural field search space summary. $w_0$ values are only for SIREN and WIRE architectures.\label{tab:nf_ablations}}
\begin{tabular}{rc}
\hline
\multicolumn{1}{c}{Knob} & Range                                \\ \hline
Activations              & Sine, Gabor, ReLU, SiLU, GELU        \\
Coordinate embedding     & Linear, SinCos, Discrete             \\
$w_0^{\text{initial}}$   & 0.1, 0.5, 1.0                        \\
$w_0^{\text{hidden}}$    & 0.5, 2.0, 10.0                       \\
Skip connections         & Yes, No                              \\
Learning rate            & $1e-2$, $5e-3$                       \\ \hline
\end{tabular}
\end{table}

An extensive grid search search was conducted evaluating every combination from \cref{tab:nf_ablations} in the $\sim1{,}000\times$ compression regime, on 12 randomly sampled density fields $\bm{f}$ from four different trajectories.
For simplicity we use PSNR of $\bm{f}$ as the selection metric. All models are trained for 10 epochs using the AdamW optimizer \citet{loshchilov2019decoupledweightdecayregularization} with a batch size of 2048.
A total of $12 \cdot 36 \:\text{(SIREN)} + 12 \cdot 18 \:\text{(WIRE)} + 12 \cdot 18 \:\text{(MLP)}=864$ neural fields were trained for this ablation.
The results from Tables \ref{tab:mlp_combinations}, \ref{tab:siren_combinations}, and \ref{tab:wire_combinations} suggest that MLP with SiLU activation, skip connections and discrete index embedding is the most performant setup, as well as the fastest and easiest to tune.

\begin{table}[h]
\centering
\renewcommand{\arraystretch}{0.9}
\caption{MLP grid search combinations.\label{tab:mlp_combinations}}
\begin{tabular}{lccccl}
\toprule
Activation & Embedding & Skip & Learning rate & $\bm{f}$ PSNR \\
\midrule
SiLU & Discrete & Yes & $5e{-3}$ & 40.53 \\
GELU & Discrete & Yes & $5e{-3}$ & 40.12 \\
SiLU & Discrete & No & $5e{-3}$ & 40.11 \\
GELU & Discrete & No & $5e{-3}$ & 39.96 \\
ReLU & Discrete & Yes & $5e{-3}$ & 39.24 \\
ReLU & Discrete & No & $5e{-3}$ & 38.83 \\
GELU & Linear & No & $5e{-3}$ & 37.06 \\
SiLU & SinCos & No & $5e{-3}$ & 36.88 \\
GELU & SinCos & No & $5e{-3}$ & 36.78 \\
GELU & Linear & Yes & $5e{-3}$ & 36.7 \\
SiLU & Linear & No & $5e{-3}$ & 36.47 \\
GELU & SinCos & Yes & $5e{-3}$ & 36.44 \\
SiLU & Linear & Yes & $5e{-3}$ & 36.09 \\
SiLU & SinCos & Yes & $5e{-3}$ & 35.18 \\
ReLU & SinCos & Yes & $5e{-3}$ & 35.1 \\
ReLU & SinCos & No & $5e{-3}$ & 34.68 \\
ReLU & Linear & No & $5e{-3}$ & 34.45 \\
ReLU & Linear & Yes & $5e{-3}$ & 34.4 \\
\bottomrule
\end{tabular}
\end{table}

\begin{table}[h]
\centering
\renewcommand{\arraystretch}{1.0}
\caption{WIRE grid search combinations.\label{tab:wire_combinations}}
\begin{tabular}{lccccl}
\toprule
Embedding & $w_0^{\text{initial}}$ & $w_0^{\text{hidden}}$ & Learning rate & $\bm{f}$ PSNR \\
\midrule
Discrete & 0.5 & 2.0 & $1e{-2}$ & 29.33 \\
Discrete & 0.1 & 2.0 & $1e{-2}$ & 27.96 \\
Discrete & 0.5 & 0.5 & $1e{-2}$ & 27.9 \\
Discrete & 0.1 & 0.5 & $1e{-2}$ & 27.83 \\
Linear & 0.1 & 2.0 & $1e{-2}$ & 24.16 \\
Linear & 0.1 & 5.0 & $1e{-2}$ & 24.16 \\
Linear & 0.1 & 0.5 & $1e{-2}$ & 24.16 \\
Linear & 0.5 & 0.5 & $1e{-2}$ & 24.16 \\
Linear & 0.5 & 2.0 & $1e{-2}$ & 24.16 \\
Linear & 0.5 & 5.0 & $1e{-2}$ & 24.16 \\
Discrete & 1.0 & 0.5 & $1e{-2}$ & 7.65 \\
Discrete & 1.0 & 2.0 & $1e{-2}$ & 7.34 \\
Linear & 1.0 & 0.5 & $1e{-2}$ & 6.04 \\
Linear & 1.0 & 2.0 & $1e{-2}$ & 6.04 \\
Linear & 1.0 & 5.0 & $1e{-2}$ & 6.04 \\
Discrete & 0.1 & 5.0 & $1e{-2}$ & nan \\
Discrete & 0.5 & 5.0 & $1e{-2}$ & nan \\
Discrete & 1.0 & 5.0 & $1e{-2}$ & nan \\
\bottomrule
\end{tabular}
\end{table}

\begin{table}[h]
\centering
\renewcommand{\arraystretch}{0.9}
\caption{SIREN grid search combinations.\label{tab:siren_combinations}}
\begin{tabular}{lccccl}
\toprule
Embedding & $w_0^{\text{initial}}$ & $w_0^{\text{hidden}}$ & Skip & Learning rate & $\bm{f}$ PSNR \\
\midrule
Discrete & 0.1 & 0.5 & Yes & $5e{-3}$ & 40.48 \\
Discrete & 0.5 & 0.5 & Yes & $5e{-3}$ & 40.34 \\
Discrete & 0.5 & 0.5 & No & $5e{-3}$ & 40.04 \\
Discrete & 0.1 & 0.5 & No & $5e{-3}$ & 39.97 \\
SinCos & 0.5 & 2.0 & Yes & $5e{-3}$ & 38.24 \\
SinCos & 0.1 & 2.0 & Yes & $5e{-3}$ & 38.19 \\
SinCos & 0.5 & 0.5 & No & $5e{-3}$ & 37.22 \\
SinCos & 0.1 & 0.5 & No & $5e{-3}$ & 37.2 \\
SinCos & 0.1 & 0.5 & Yes & $5e{-3}$ & 36.23 \\
SinCos & 0.5 & 0.5 & Yes & $5e{-3}$ & 36.23 \\
SinCos & 0.1 & 2.0 & No & $5e{-3}$ & 32.58 \\
Discrete & 0.1 & 2.0 & No & $5e{-3}$ & 29.41 \\
SinCos & 0.1 & 5.0 & Yes & $5e{-3}$ & 24.16 \\
SinCos & 0.1 & 5.0 & No & $5e{-3}$ & 24.16 \\
Discrete & 0.1 & 5.0 & No & $5e{-3}$ & 24.16 \\
Discrete & 0.1 & 2.0 & Yes & $5e{-3}$ & 24.16 \\
Discrete & 0.5 & 2.0 & Yes & $5e{-3}$ & 24.16 \\
Discrete & 0.1 & 5.0 & Yes & $5e{-3}$ & 24.16 \\
Discrete & 1.0 & 0.5 & Yes & $5e{-3}$ & 10.1 \\
Discrete & 1.0 & 0.5 & No & $5e{-3}$ & 10.03 \\
SinCos & 1.0 & 2.0 & Yes & $5e{-3}$ & 9.57 \\
SinCos & 1.0 & 0.5 & No & $5e{-3}$ & 9.29 \\
SinCos & 1.0 & 0.5 & Yes & $5e{-3}$ & 9.04 \\
SinCos & 1.0 & 2.0 & No & $5e{-3}$ & 8.74 \\
SinCos & 0.5 & 2.0 & No & $5e{-3}$ & 8.43 \\
Discrete & 1.0 & 2.0 & No & $5e{-3}$ & 6.99 \\
Discrete & 0.5 & 2.0 & No & $5e{-3}$ & 6.94 \\
Discrete & 1.0 & 2.0 & Yes & $5e{-3}$ & 6.08 \\
SinCos & 1.0 & 5.0 & Yes & $5e{-3}$ & 6.04 \\
SinCos & 0.5 & 5.0 & Yes & $5e{-3}$ & 6.04 \\
Discrete & 0.5 & 5.0 & No & $5e{-3}$ & 6.04 \\
SinCos & 1.0 & 5.0 & No & $5e{-3}$ & 6.04 \\
Discrete & 1.0 & 5.0 & No & $5e{-3}$ & 6.04 \\
SinCos & 0.5 & 5.0 & No & $5e{-3}$ & 6.04 \\
Discrete & 1.0 & 5.0 & Yes & $5e{-3}$ & 6.04 \\
Discrete & 0.5 & 5.0 & Yes & $5e{-3}$ & 6.04 \\
\bottomrule
\end{tabular}
\end{table}

\clearpage

\subsection{Extra results}
\label{app:extra_results}

\begin{table}[h]
\renewcommand{\arraystretch}{1.1}
\scriptsize
\caption{Missing metrics from \cref{tab:results}.
Evaluation on 420 total $\bm{f}$s (20 different turbulent trajectories, 21 random time snapshots), sampled in the statistically steady phase, same setup as \cref{tab:results}. Errors in data space.
\label{tab:missing_results}}
\centering
\resizebox{\textwidth}{!}{%
\begin{tabular}{l|cc|c|cccc}
\toprule
         &  \multicolumn{2}{c}{Compression $\bm{f}$} & \multicolumn{1}{c}{Integrals $\bm{\phi}$} & \multicolumn{4}{c}{Turbulence $Q^{\text{spec}}, k_y^{\text{spec}}$}                                        \\ \midrule
         & $\text{L1}(\bm{f})$ $\downarrow$ & BPP & $\text{L1}(\bm{\phi})$ $\downarrow$ & $\text{PC}(\overline{k_y^{\text{spec}}})$ $\uparrow$ & $\text{PC}(\overline{Q^{\text{spec}}})$ $\uparrow$ & $\text{L1}(\overline{k_y^{\text{spec}}})$ $\uparrow$ & $\text{L1}(\overline{Q^{\text{spec}}})$ $\uparrow$ \\
\midrule
ZFP      & 0.72$_{\pm 0.19}$ & 0.036 & $1.11\times10^{3}{}_{\pm 1.06\times10^{3}}$ & 0.9287$_{\pm 0.10}$ & -0.0053$_{\pm 0.73}$ & $4.37\times10^{5}{}_{\pm 8.59\times10^{5}}$ & 107.51$_{\pm 40.90}$ \\
Wavelet  & 0.50$_{\pm 0.12}$ & 0.034 & $7.49\times10^{2}{}_{\pm 8.54\times10^{2}}$ & 0.9291$_{\pm 0.10}$ & -0.9628$_{\pm 0.05}$ & $3.90\times10^{5}{}_{\pm 8.27\times10^{5}}$ & 107.74$_{\pm 40.99}$ \\
PCA      & 0.52$_{\pm 0.14}$ & 0.031 & $4.46\times10^{2}{}_{\pm 4.19\times10^{2}}$ & 0.9290$_{\pm 0.10}$ & 0.8705$_{\pm 0.38}$ & $9.95\times10^{4}{}_{\pm 1.56\times10^{5}}$ & 67.45$_{\pm 24.02}$ \\
JPEG2000 & 0.51$_{\pm 0.13}$ & 0.032 & $1.78\times10^{3}{}_{\pm 1.52\times10^{3}}$ & 0.9280$_{\pm 0.10}$ & -0.0363$_{\pm 0.82}$ & $8.63\times10^{5}{}_{\pm 1.27\times10^{6}}$ & 103.91$_{\pm 36.75}$ \\\midrule
VAPOR & 0.88$_{\pm 0.25}$ & 0.5 & $7.80\times10^{2}{}_{\pm 6.60\times10^{2}}$ & 0.9280$_{\pm 0.10}$ & 0.9804$_{\pm 0.01}$ & $1.62\times10^{5}{}_{\pm 2.80\times10^{5}}$ & 91.95$_{\pm 40.39}$ \\\midrule
NF       & 0.33$_{\pm 0.09}$ & 0.027 & $1.37\times10^{2}{}_{\pm 1.26\times10^{2}}$ & 0.9392$_{\pm 0.08}$ & 0.9735$_{\pm 0.02}$ & $6.07\times10^{3}{}_{\pm 1.01\times10^{4}}$ & 61.50$_{\pm 14.96}$ \\
PINC-NF  & 0.37$_{\pm 0.08}$ & 0.027 & 26.52$_{\pm 24.35}$ & 0.9870$_{\pm 0.02}$ & 0.9797$_{\pm 0.02}$ & $2.37\times10^{2}{}_{\pm 3.92\times10^{2}}$ & 49.48$_{\pm 15.88}$ \\
AE + EVA & 0.45$_{\pm 0.17}$ & 0.045 & 73.13$_{\pm 74.30}$ & 0.9419$_{\pm 0.07}$ & 0.9785$_{\pm 0.03}$ & $1.63\times10^{3}{}_{\pm 3.83\times10^{3}}$ & 48.48$_{\pm 33.26}$ \\
VQ-VAE + EVA & 0.54$_{\pm 0.20}$ & 0.0004 & 56.11$_{\pm 37.12}$ & 0.9437$_{\pm 0.06}$ & 0.7888$_{\pm 0.22}$ & $5.35\times10^{2}{}_{\pm 8.59\times10^{2}}$ & 73.47$_{\pm 46.51}$ \\
\bottomrule
\end{tabular}
}
\end{table}

\begin{figure}[h]
\centering
\includegraphics[width=\textwidth]{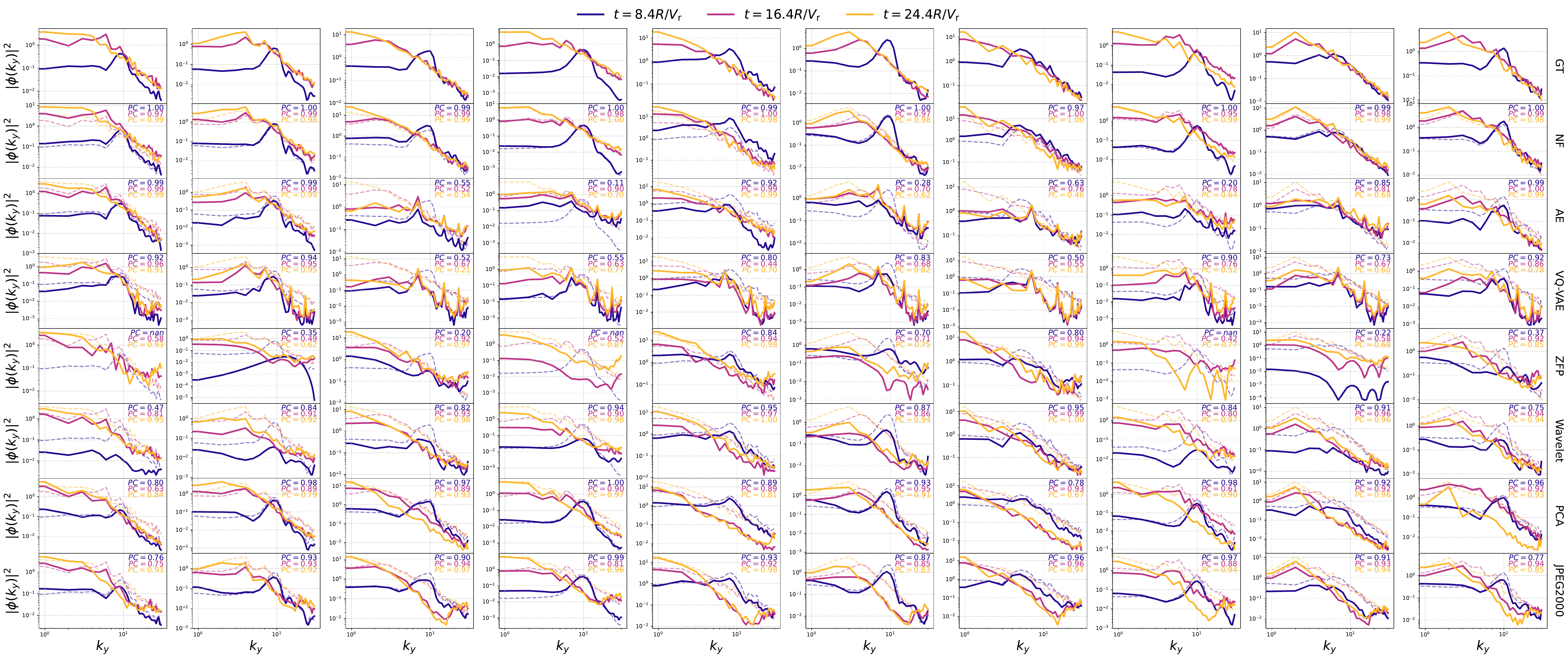}
\caption{Extra models for the energy cascade (left \cref{fig:cascade}). The three time snapshots at $[8.4, 16.4, 24.4]R/V_{\mathrm{r}}$ are specifically sampled in the transitional phase where mode growth and energy cascade happens, before reaching the statistically stable phase. Visualized as the energy transfer from higher to lower modes as turbulence develops. Columns are different trajectories, rows are compression methods, lines of varied colors are the $k_y^{\text{spec}}$ at specific timesteps, and transparent lines are respective ground truth.}
\label{fig:extra_cascade}
\end{figure}

\begin{figure}[h]
\centering
\includegraphics[width=\textwidth]{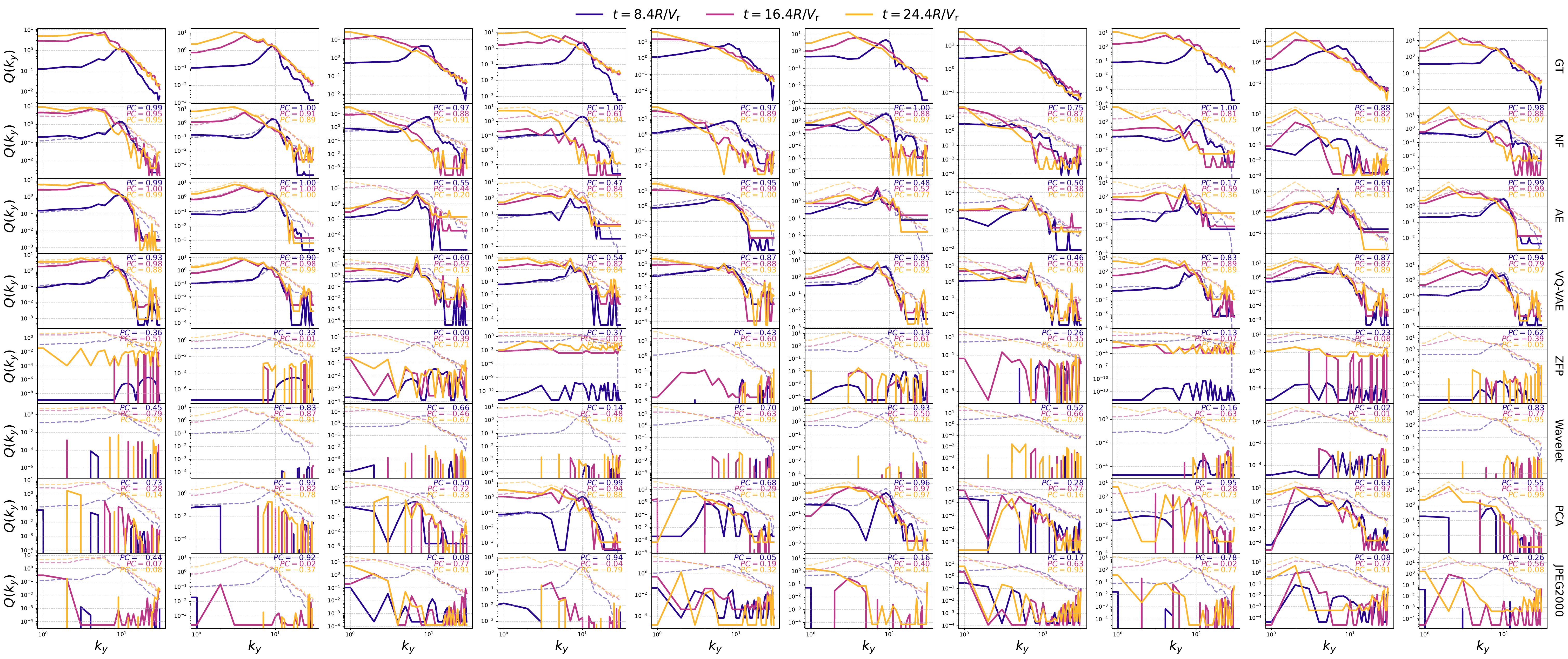}
\caption{Extra models for the $Q$ spectra (right \cref{fig:cascade}). The three time snapshots at $[8.4, 16.4, 24.4]R/V_{\mathrm{r}}$ are specifically sampled in the transitional phase where mode growth and energy cascade happens, before reaching the statistically stable phase. Visualized as the energy transfer from higher to lower modes as turbulence develops. Columns are different trajectories, rows are compression methods, lines of varied colors are the $Q^{\text{spec}}$ at specific timesteps, and transparent lines are respective ground truth.}
\label{fig:extra_qcascade}
\end{figure}

\begin{figure}[h]
\centering
\includegraphics[width=\textwidth]{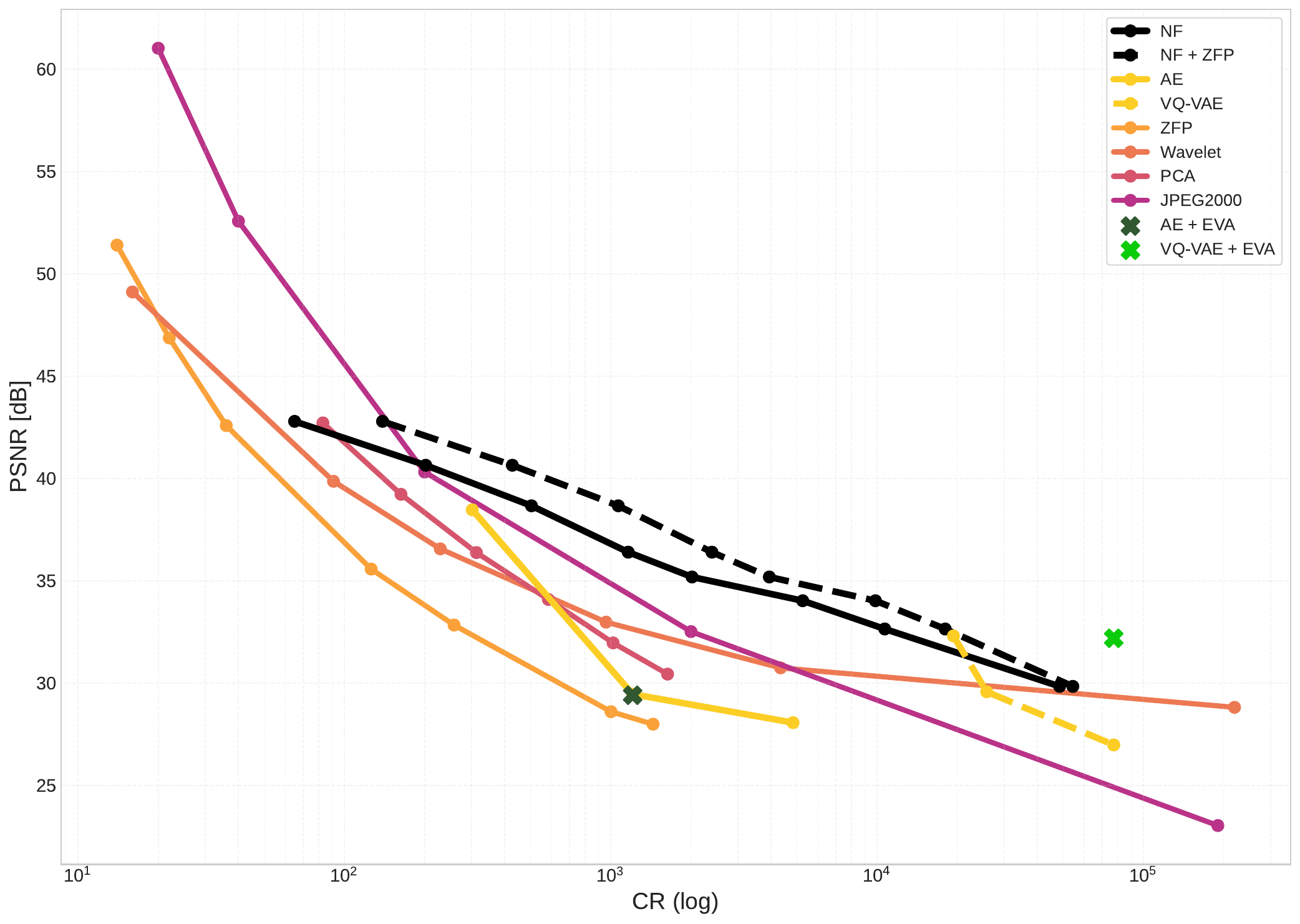}
\caption{Full PSNR scaling plot with missing curves from \cref{fig:scaling}}
\label{fig:extra_psnr_scaling}
\end{figure}

\begin{figure}[h]
\centering
\includegraphics[width=\textwidth]{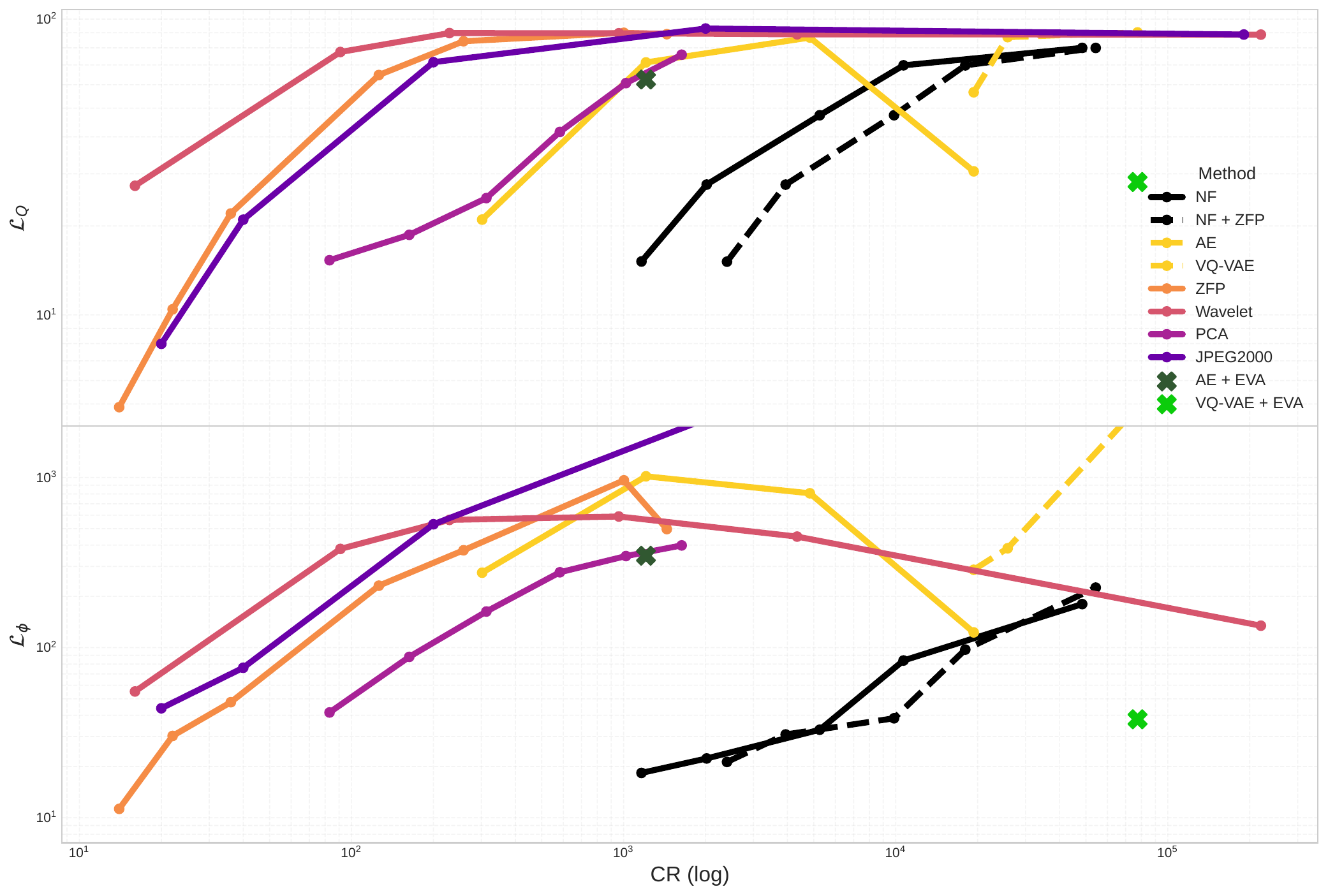}
\caption{Full physics scaling plot with missing curves from \cref{fig:phisics_scaling}}
\label{fig:extra_pinn_scaling}
\end{figure}

\begin{figure}[h]
    \centering
    \includegraphics[width=0.6\linewidth]{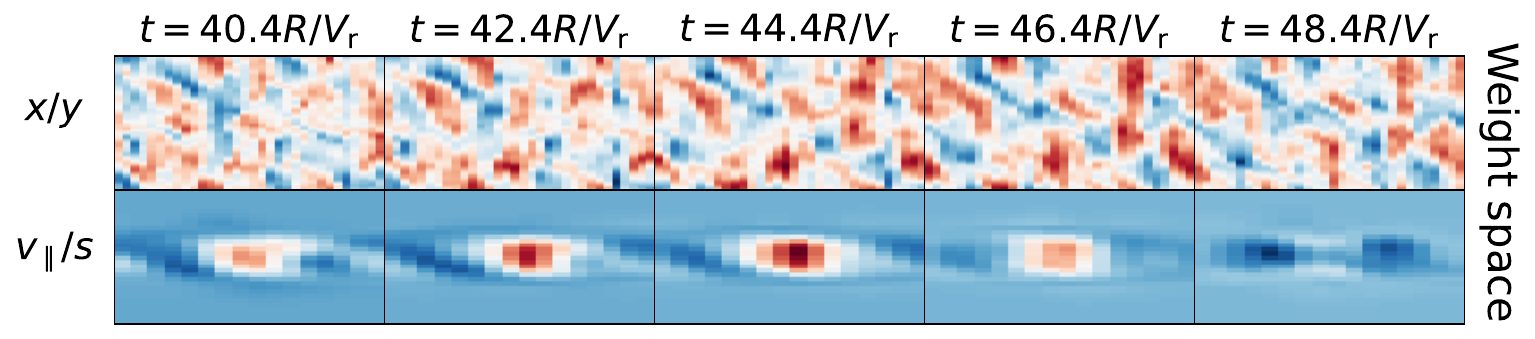}\\
    \vspace{-3.5pt}
    \includegraphics[width=0.6\linewidth]{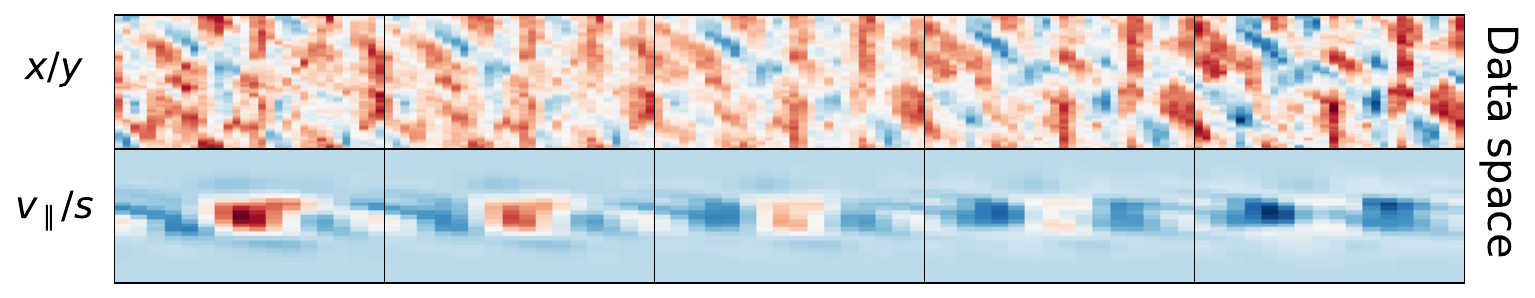}\\
    \vspace{-4.5pt}
    \includegraphics[width=0.6\linewidth]{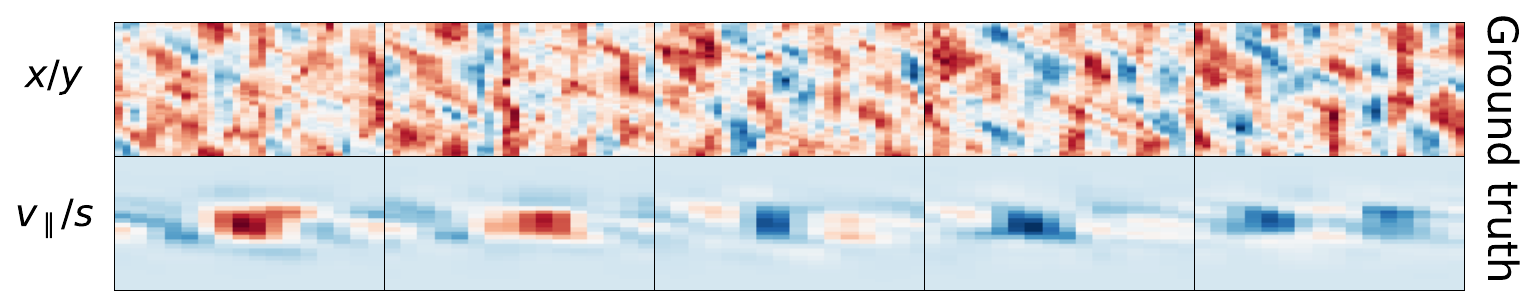}
    \caption{
    Qualitative visualization of 5D $\bm{f}$ slices interpolated over time, between the two extremes at $t=40.4R/V_{{\mathrm{{r}}}}$ and $t=48.4R/V_\mathrm{r}$.
    Aspect ratio is set to 2 and does not represent the physical one.
    }
    \label{fig:interp_slice}
\end{figure}

\begin{table}[h]
\centering
\renewcommand{\arraystretch}{1.1}
\caption{Timing details for neural and traditional compression, in seconds. GPU: single NVIDIA A40 (48GB), CPU: Intel Xeon Platinum 8168, 96 cores, 2.70GHz.\label{tab:timing}}
\begin{tabular}{lcccc}
\hline
Model & Offine compute & Compress {[}s{]} & Decompress {[}s{]} & Device \\ \hline
NF & - & 96.3 & 0.260 & GPU \\
AE & $\sim4\times 60$h $+ 28$h & 0.377 & 0.023 & GPU \\
VQ-VAE & $\sim4\times 60$h $+ 28$h & 0.425 & 0.027 & GPU \\ \hline
ZFP & - & 0.144 & 0.066 & CPU \\
Wavelet & - & 1.30 & 0.804 & CPU \\
PCA & - & 0.377 & 0.149 & CPU \\
JPEG2000 & - & 4.17 & 0.261 & CPU \\ \hline
\end{tabular}
\end{table}

\begin{figure}[t]
    \centering
    \setlength{\tabcolsep}{2pt}
    \renewcommand{\arraystretch}{1.0} 
    \begin{tabular}{cc}
        \includegraphics[width=0.45\linewidth]{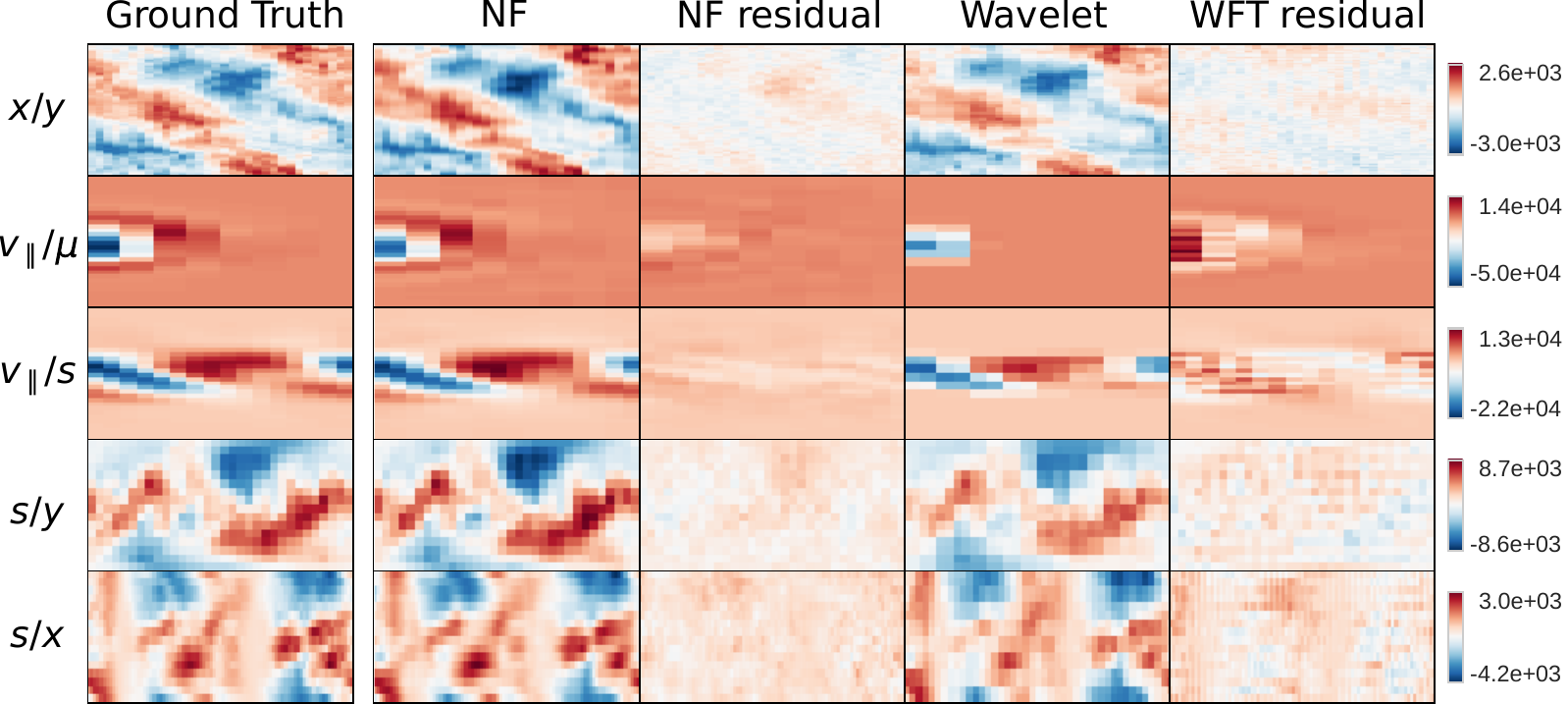} &
        \includegraphics[width=0.45\linewidth]{figs/df_recon.pdf} \\
        \includegraphics[width=0.45\linewidth]{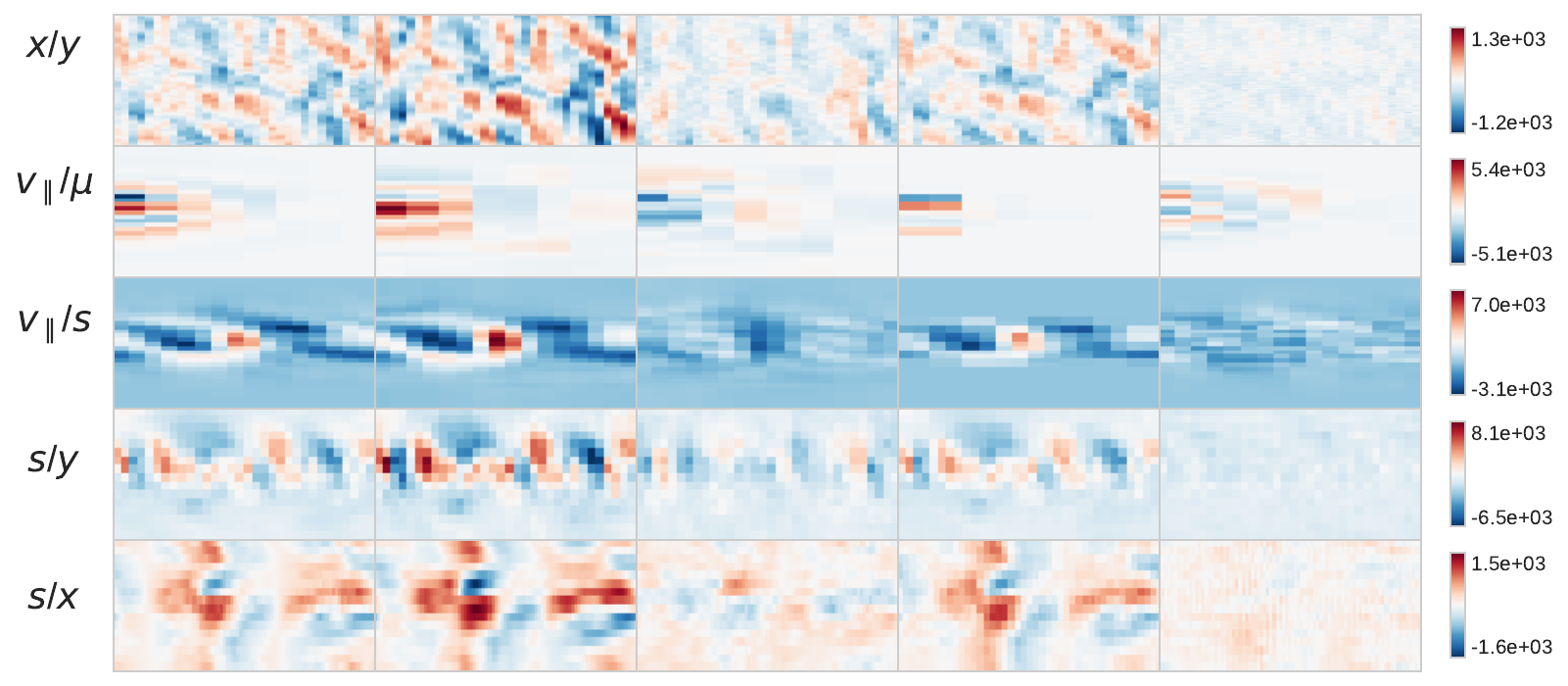} &
        \includegraphics[width=0.45\linewidth]{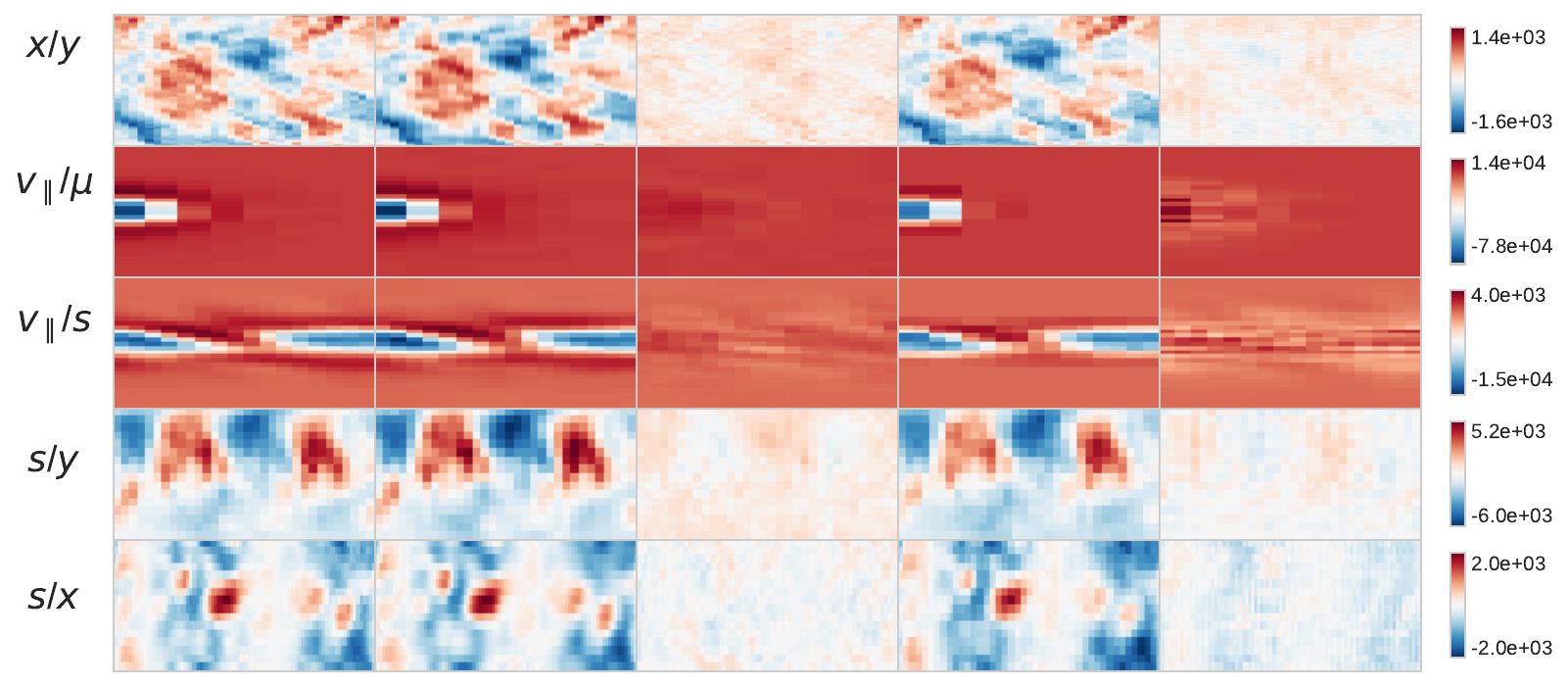} \\
        \includegraphics[width=0.45\linewidth]{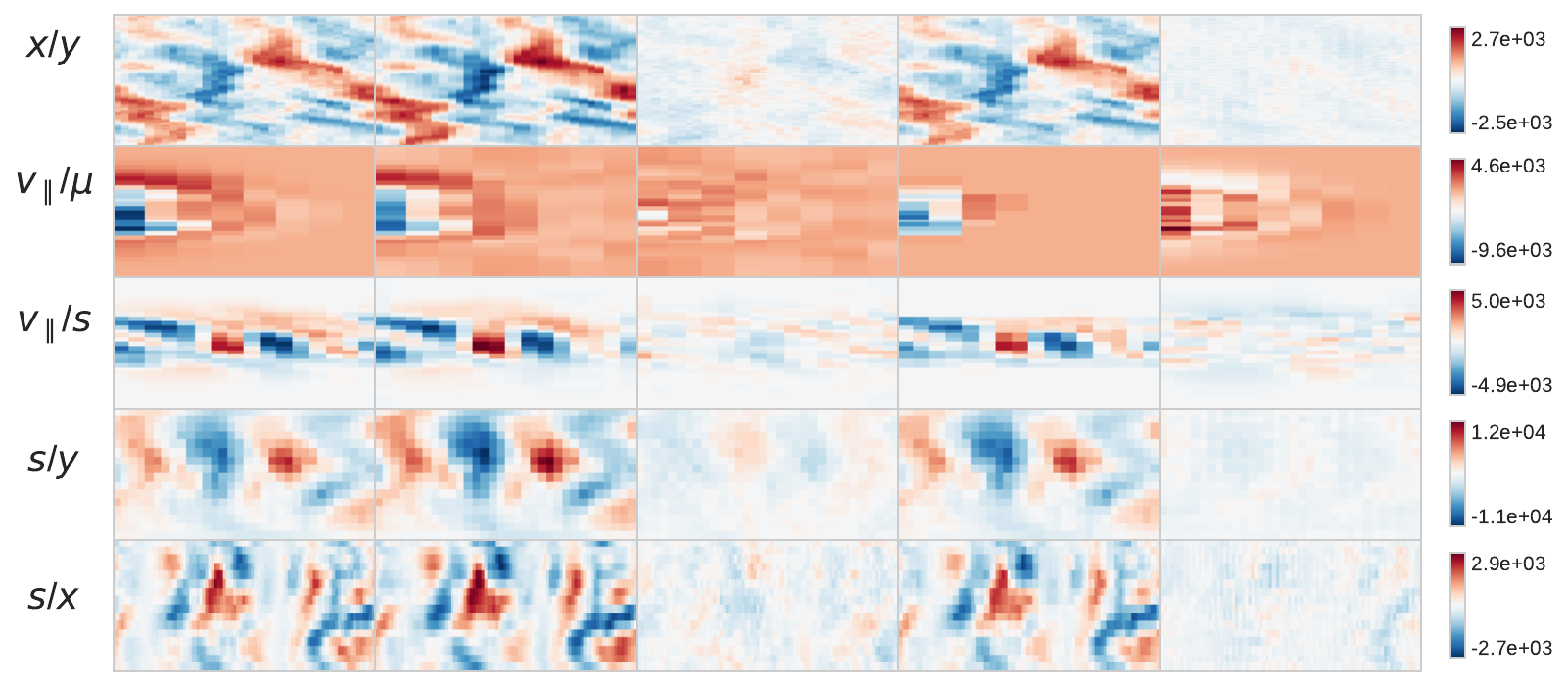} &
        \includegraphics[width=0.45\linewidth]{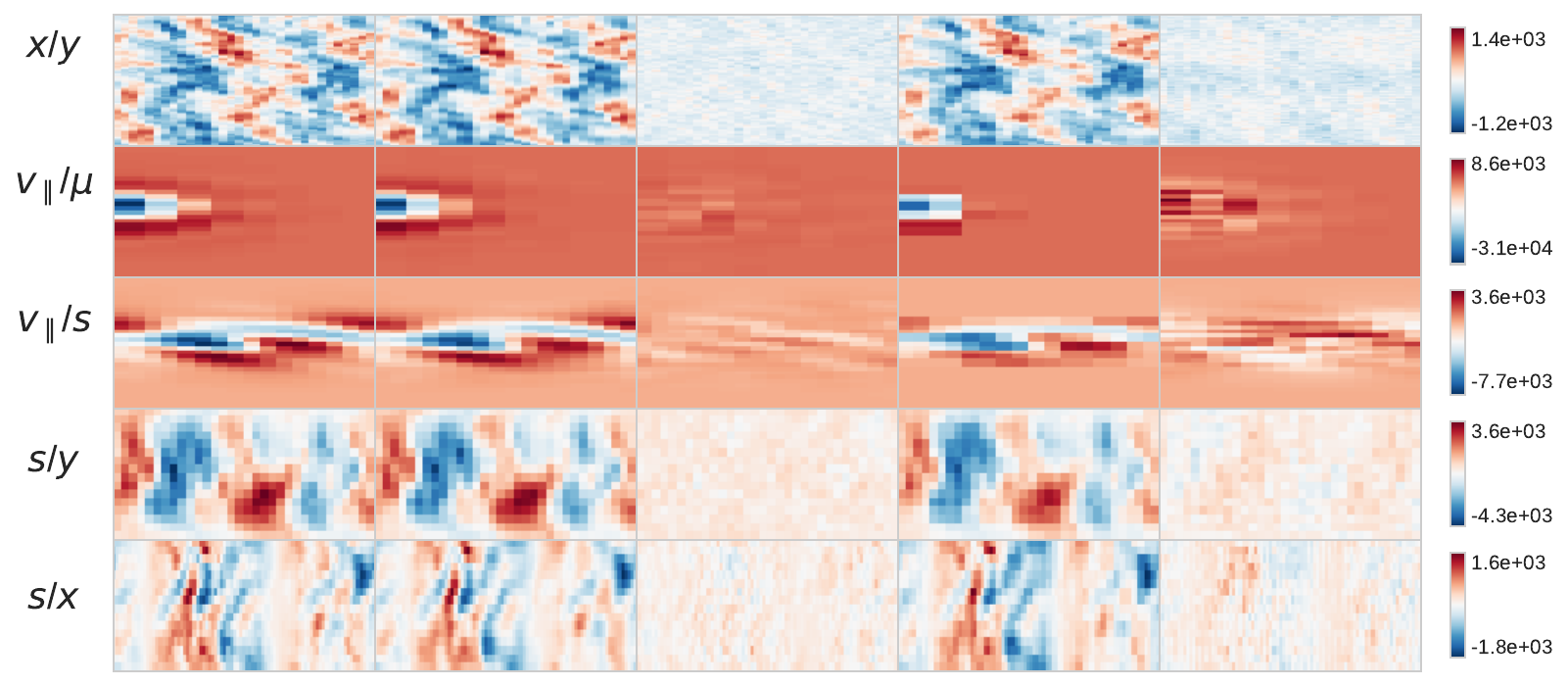} \\
        \includegraphics[width=0.45\linewidth]{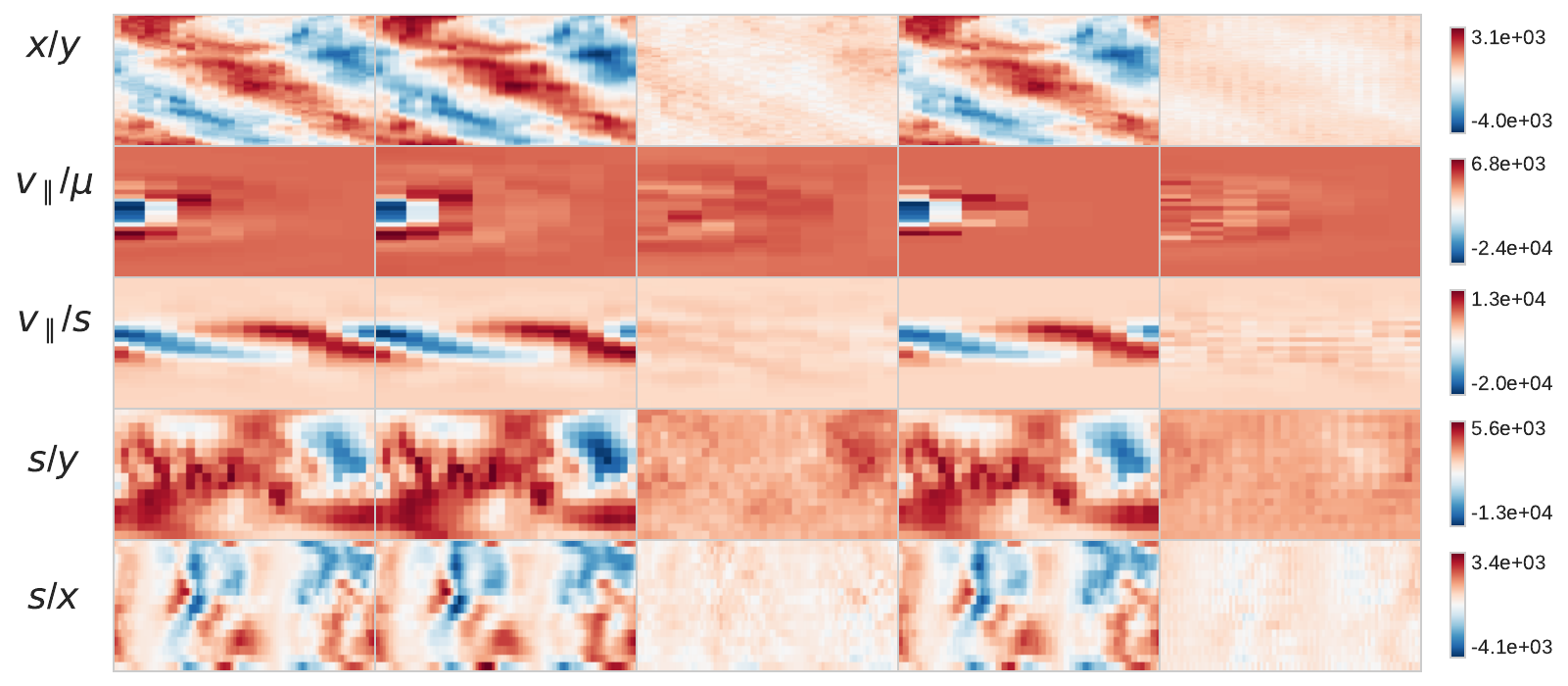} &
        \includegraphics[width=0.45\linewidth]{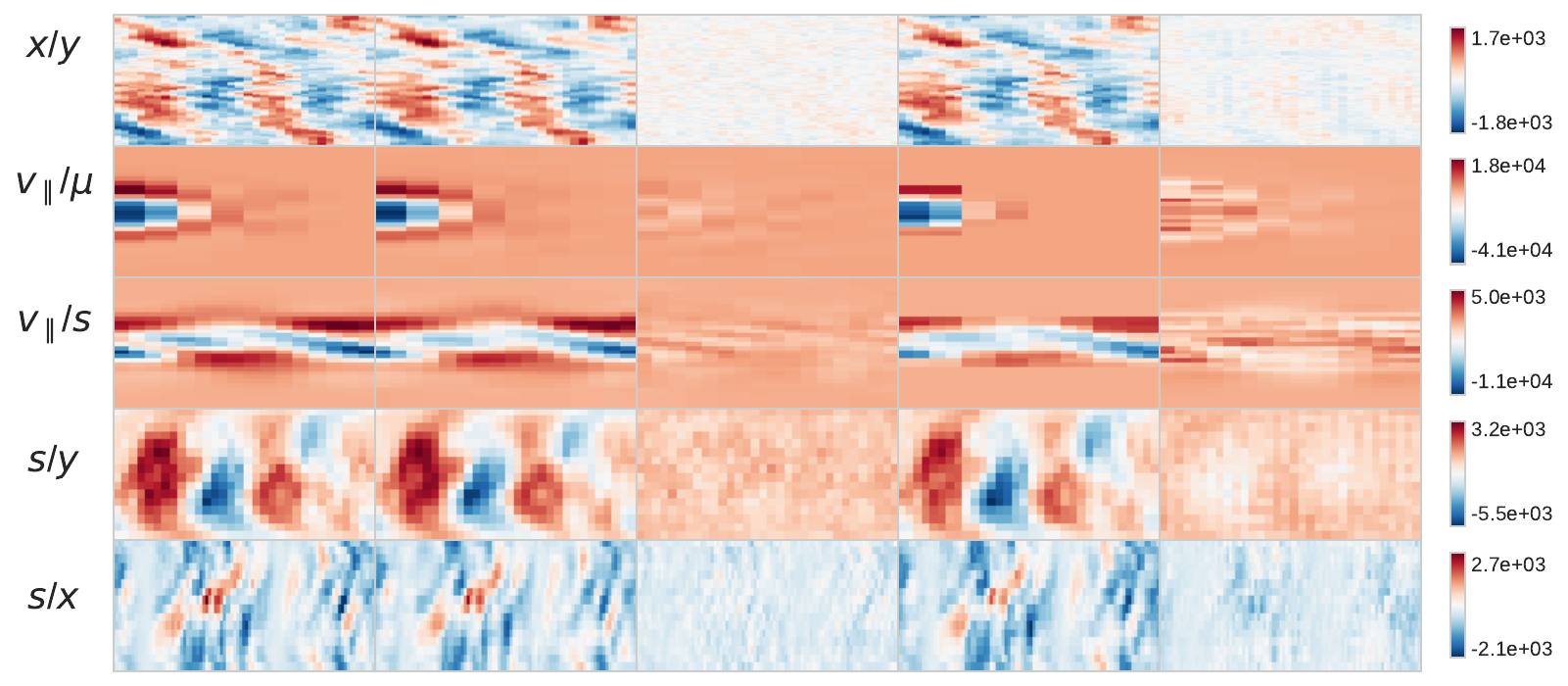} \\
        \includegraphics[width=0.45\linewidth]{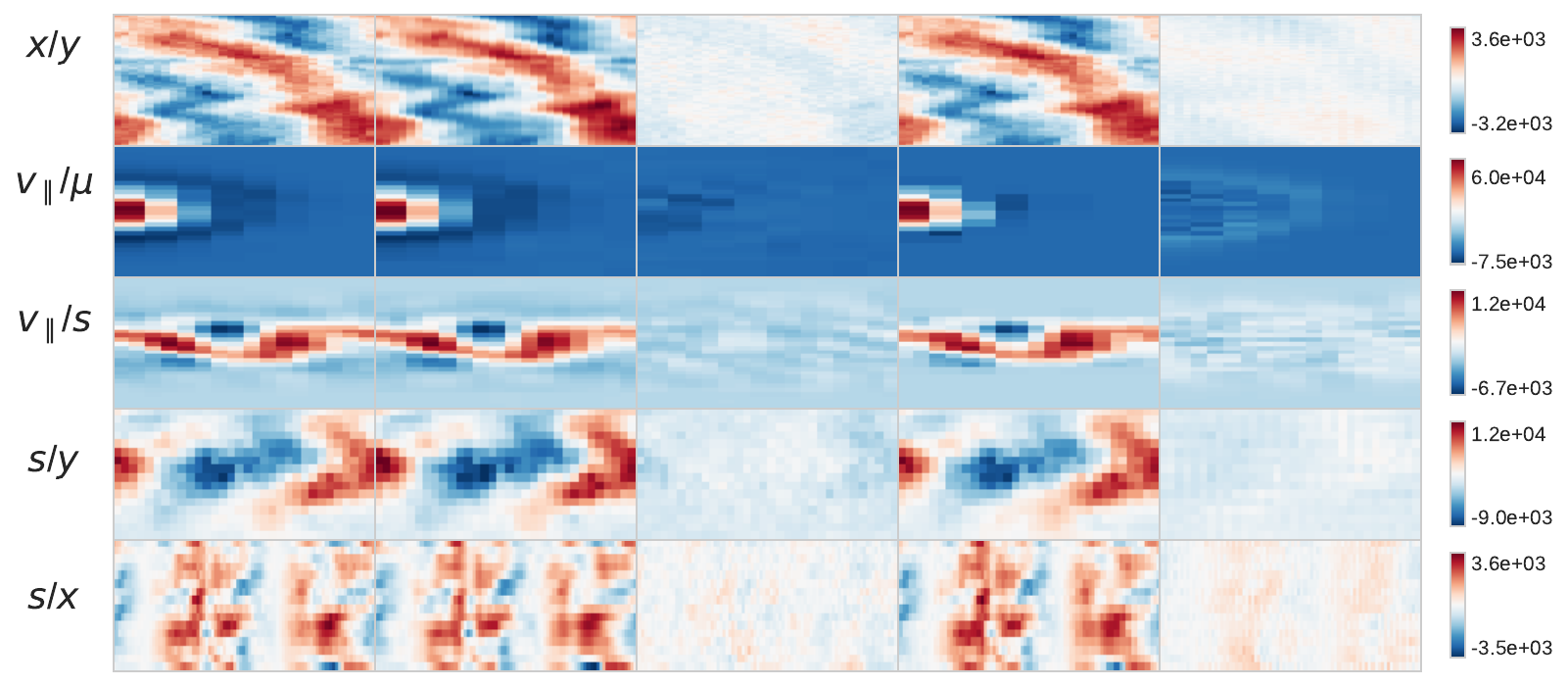} &
        \includegraphics[width=0.45\linewidth]{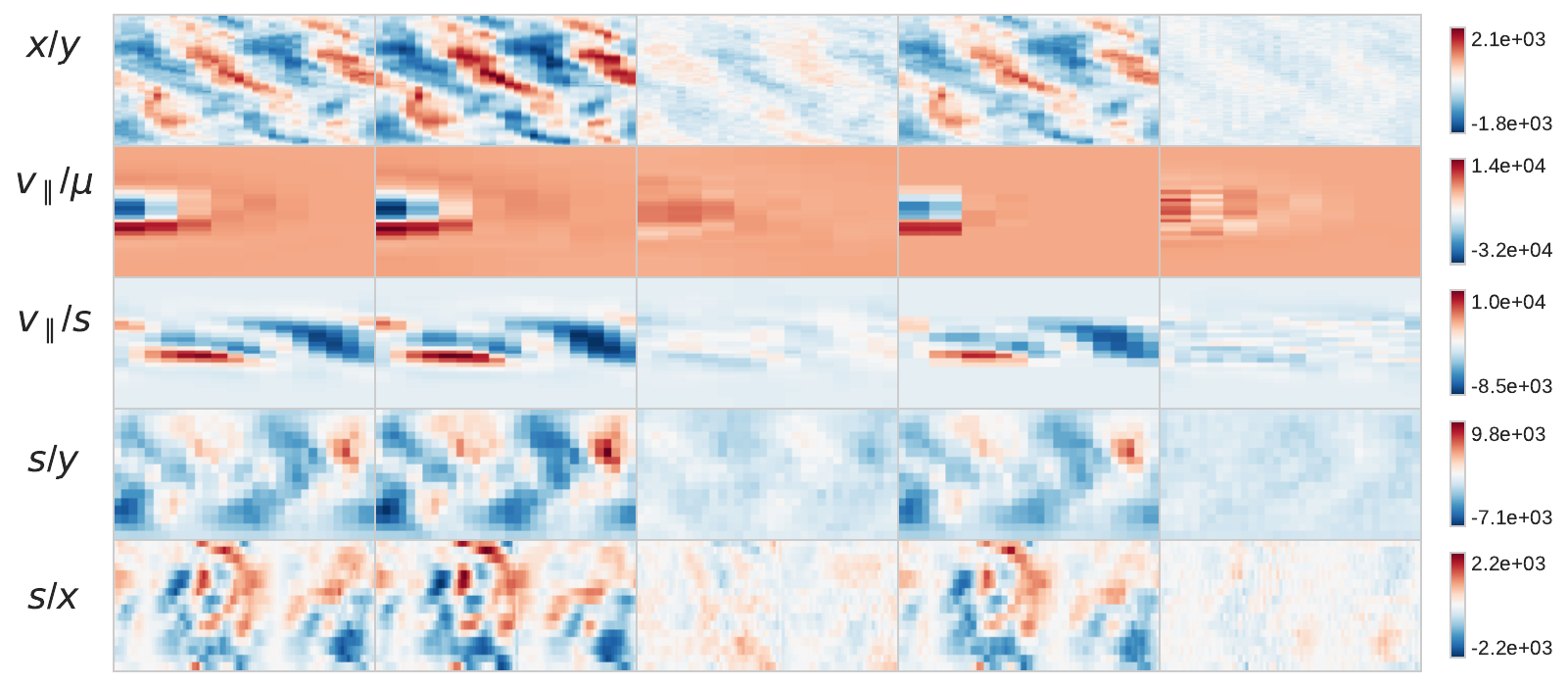} \\
        \includegraphics[width=0.45\linewidth]{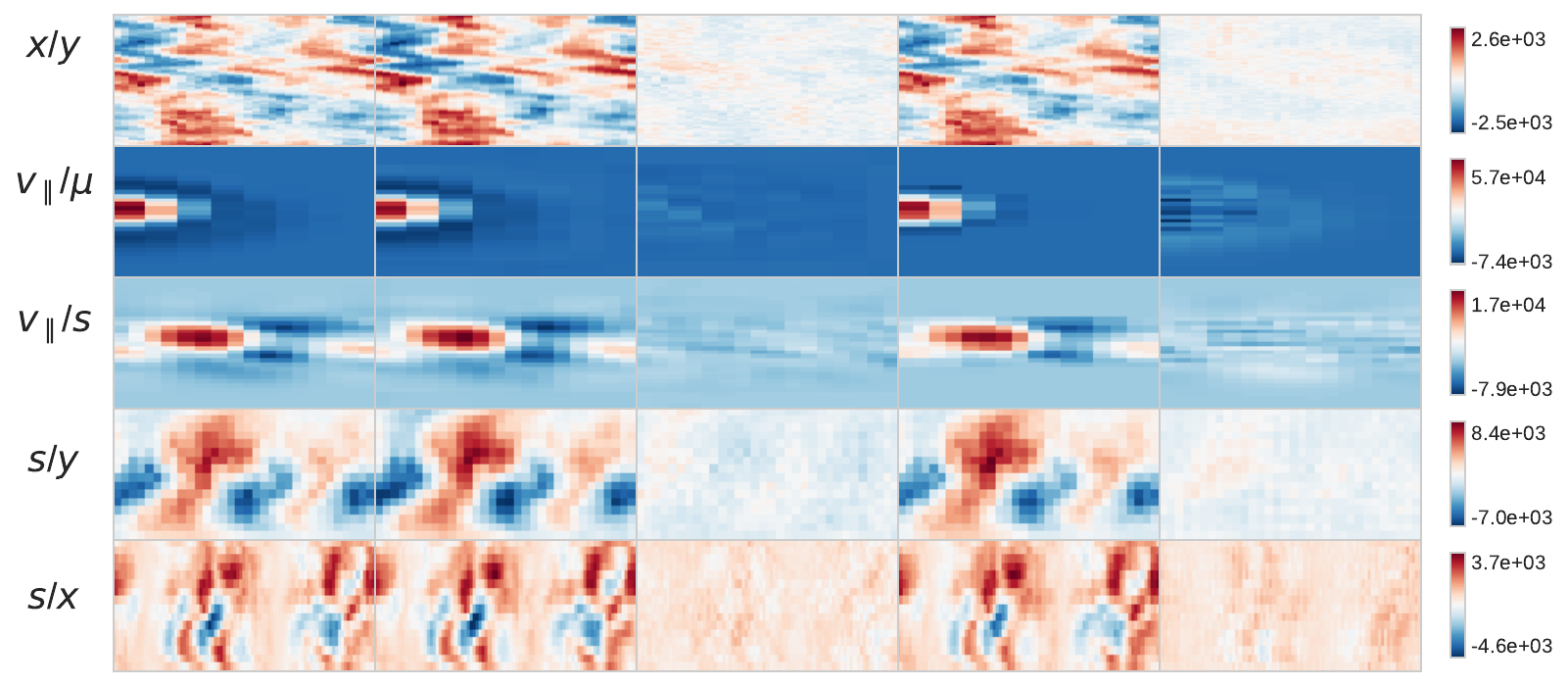} &
        \includegraphics[width=0.45\linewidth]{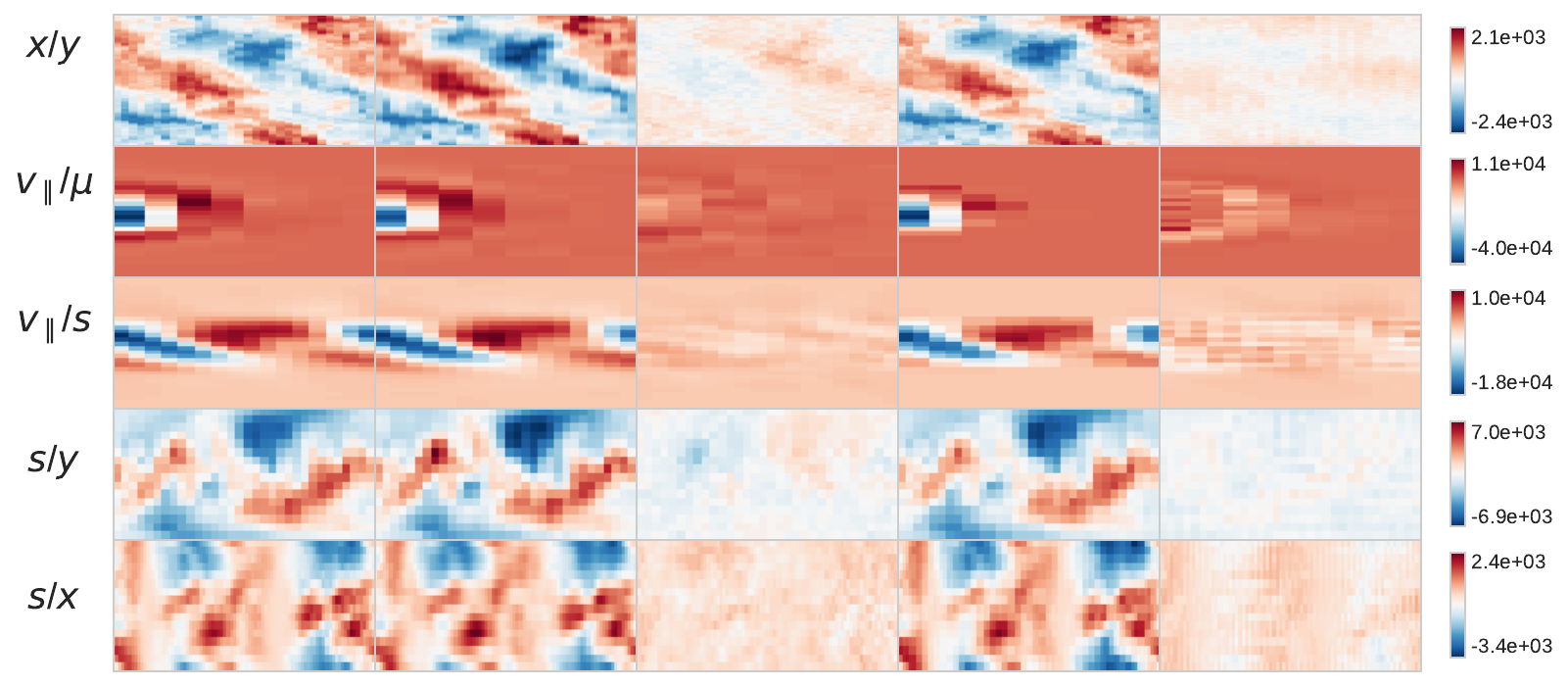} \\
    \end{tabular}
    \caption{
    Extra reconstructions for the 5D density function $\bm{f}$. $\text{CR}=\sim1{,}000\times$.
    Each row is a different trajectory at timestep $176.4R/V_\mathrm{r}$. Columns match \cref{fig:extra_phi}.
    Aspect ratio is set to 2 and does not represent the physical one.
    }
    \label{fig:extra_df}
\end{figure}

\begin{figure}[t]
    \centering
    \setlength{\tabcolsep}{2pt}
    \renewcommand{\arraystretch}{1.0}
    \begin{tabular}{cc}
        \includegraphics[width=0.45\linewidth]{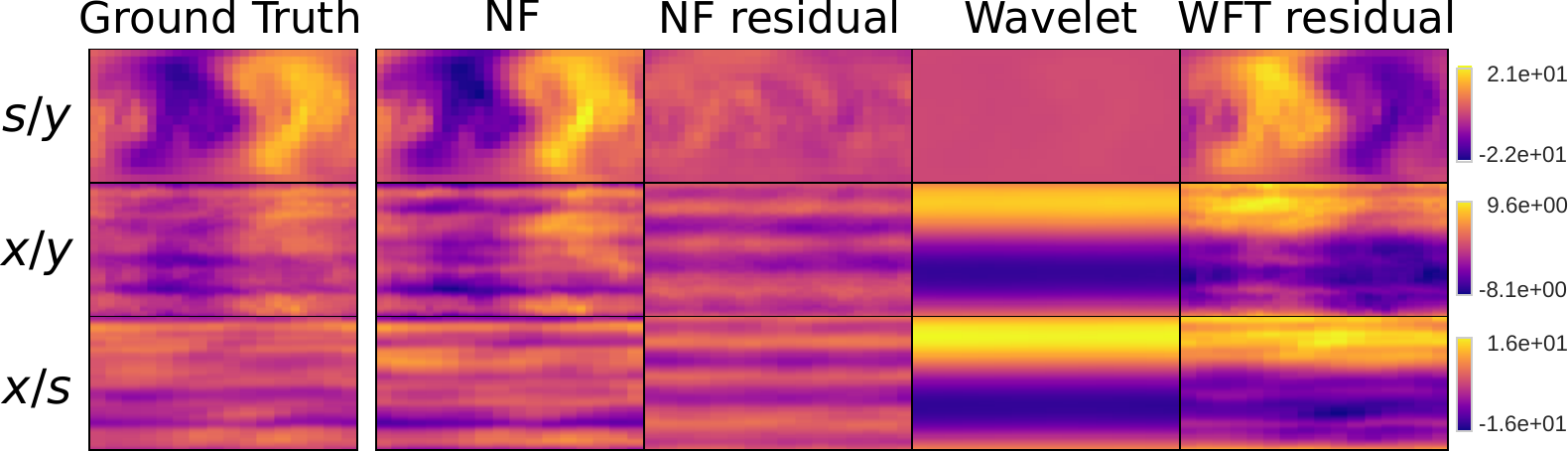} &
        \includegraphics[width=0.45\linewidth]{figs/phi_recon.pdf} \\
        \includegraphics[width=0.45\linewidth]{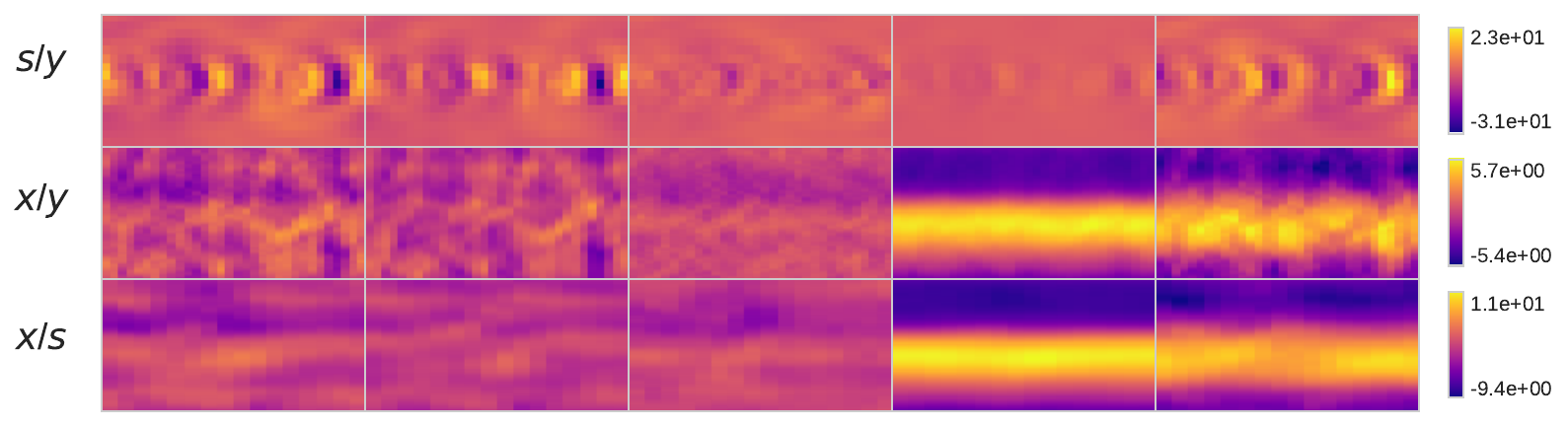} &
        \includegraphics[width=0.45\linewidth]{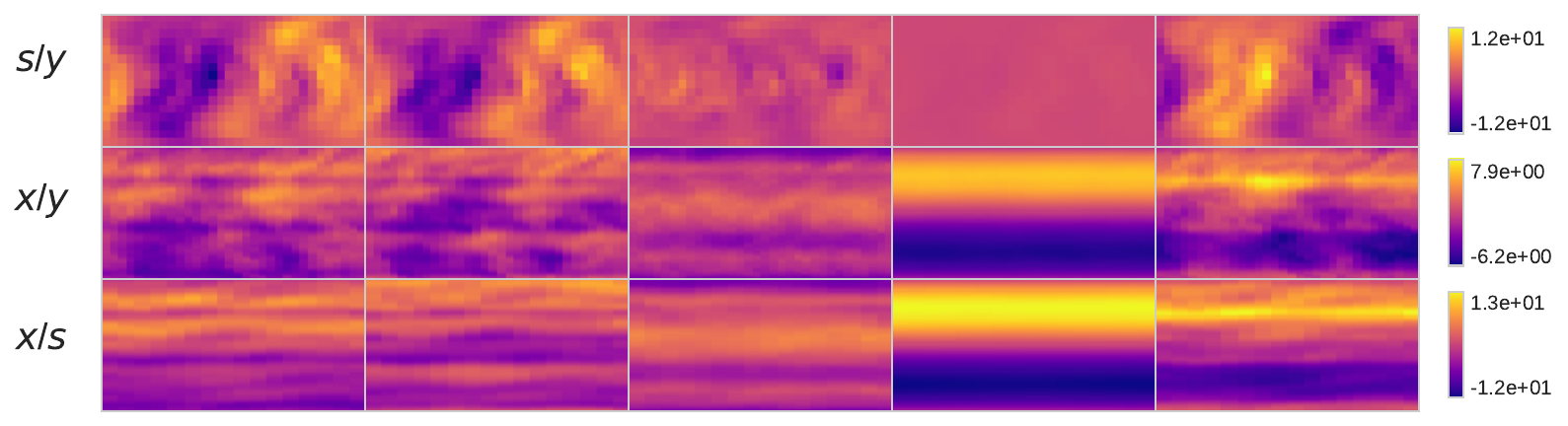} \\
        \includegraphics[width=0.45\linewidth]{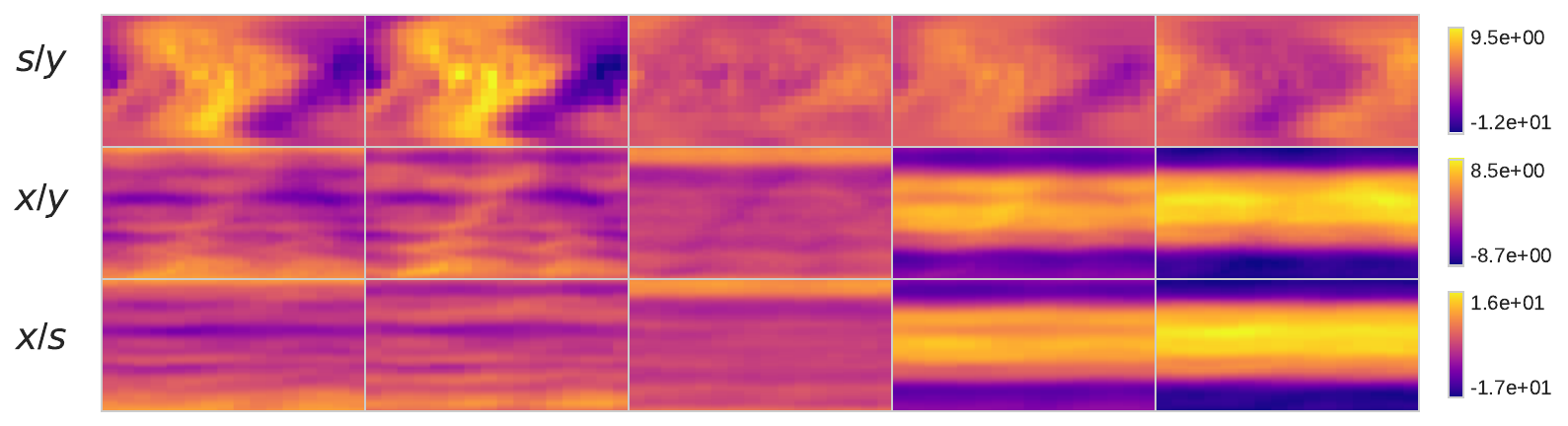} &
        \includegraphics[width=0.45\linewidth]{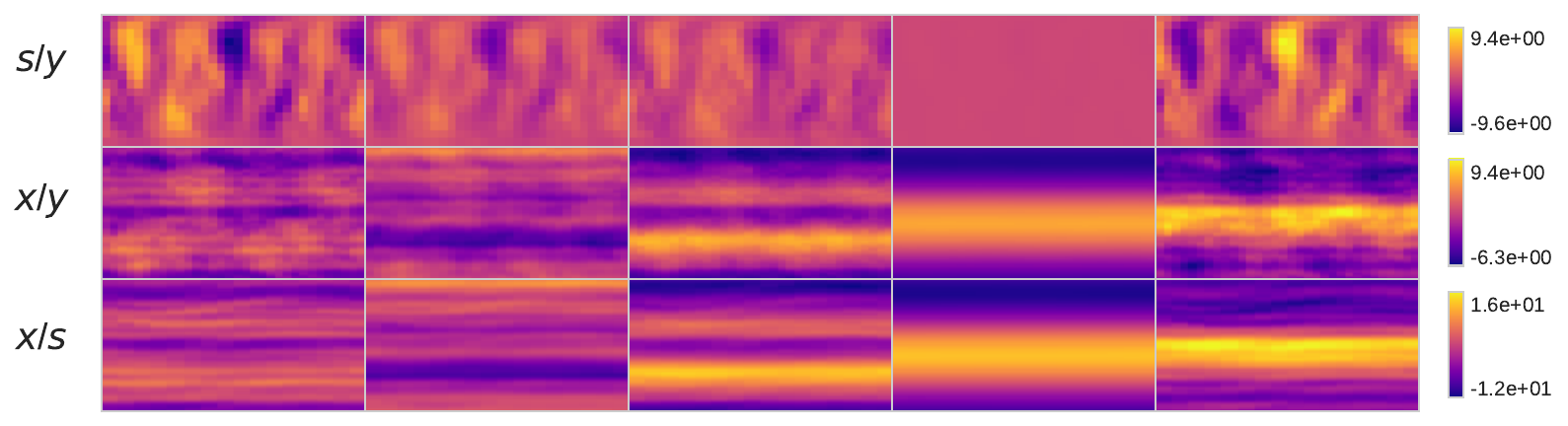} \\
        \includegraphics[width=0.45\linewidth]{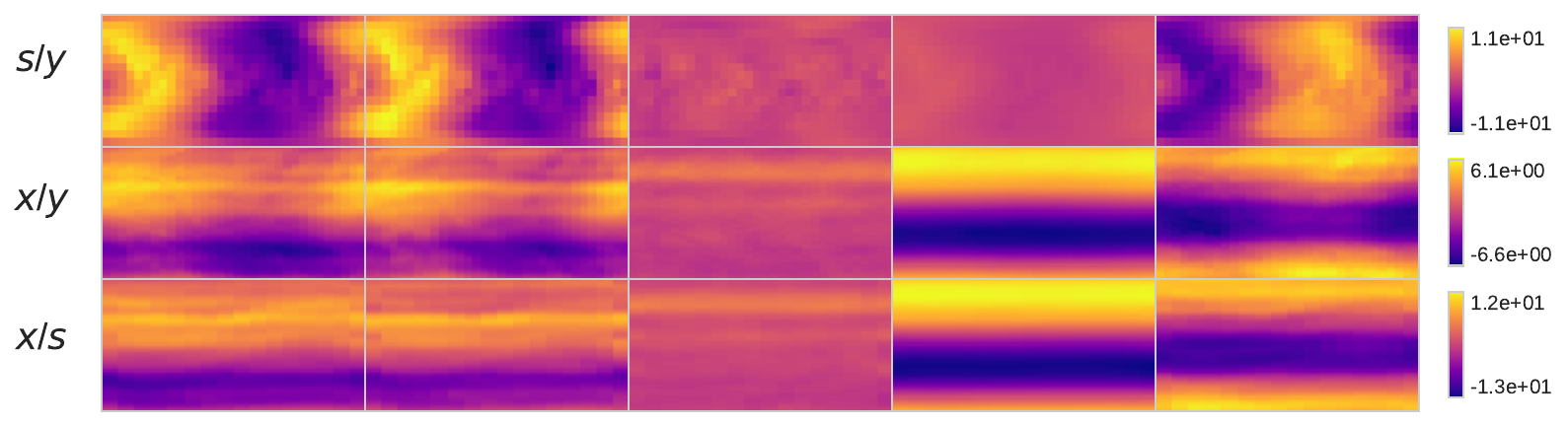} &
        \includegraphics[width=0.45\linewidth]{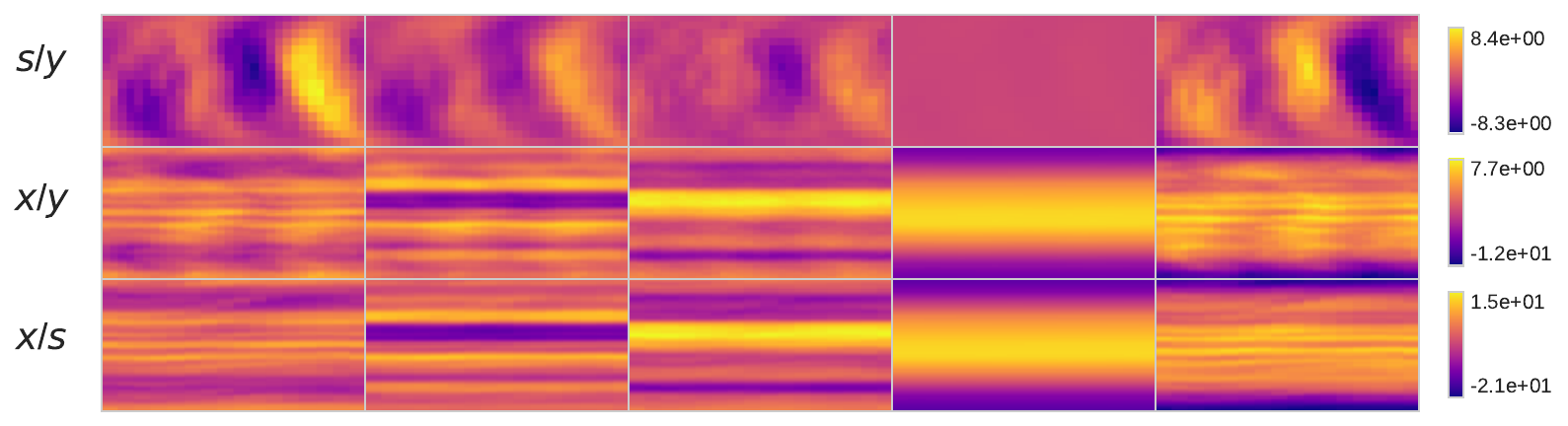} \\
        \includegraphics[width=0.45\linewidth]{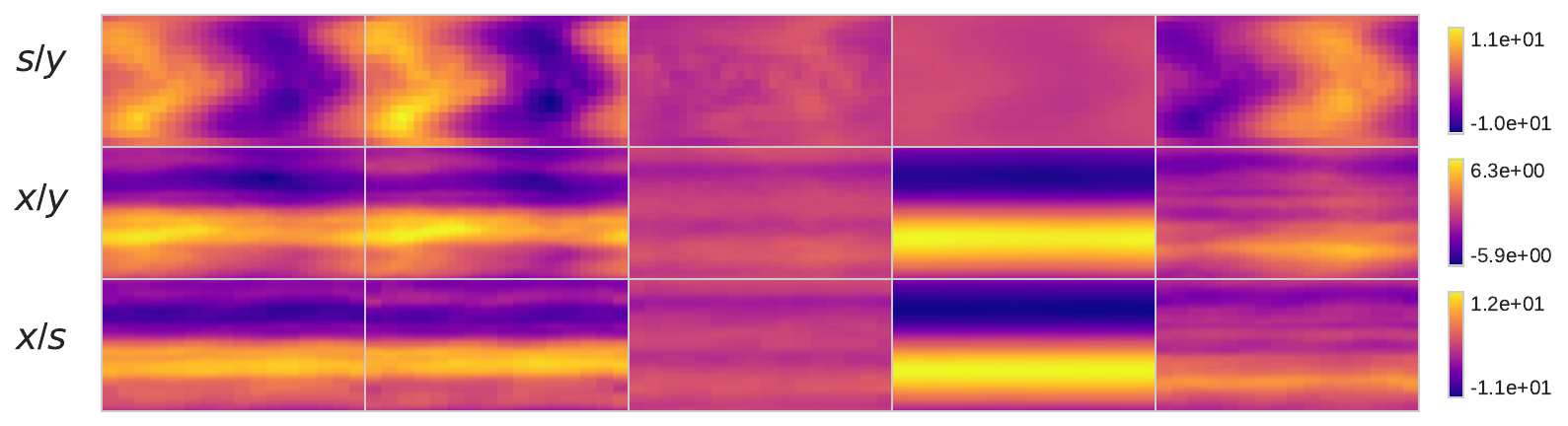} &
        \includegraphics[width=0.45\linewidth]{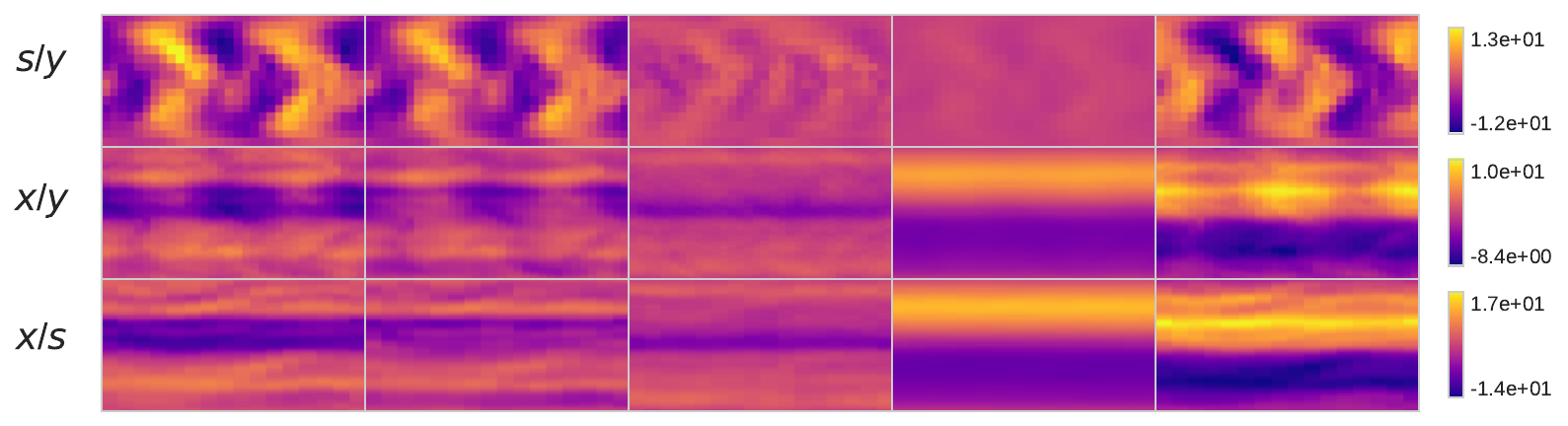} \\
        \includegraphics[width=0.45\linewidth]{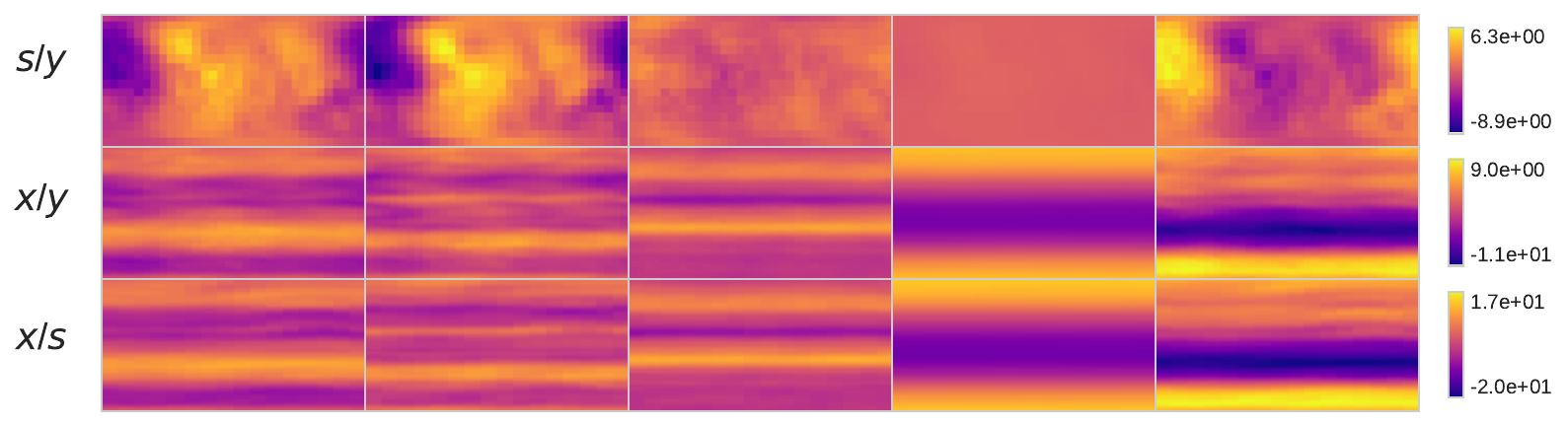} &
        \includegraphics[width=0.45\linewidth]{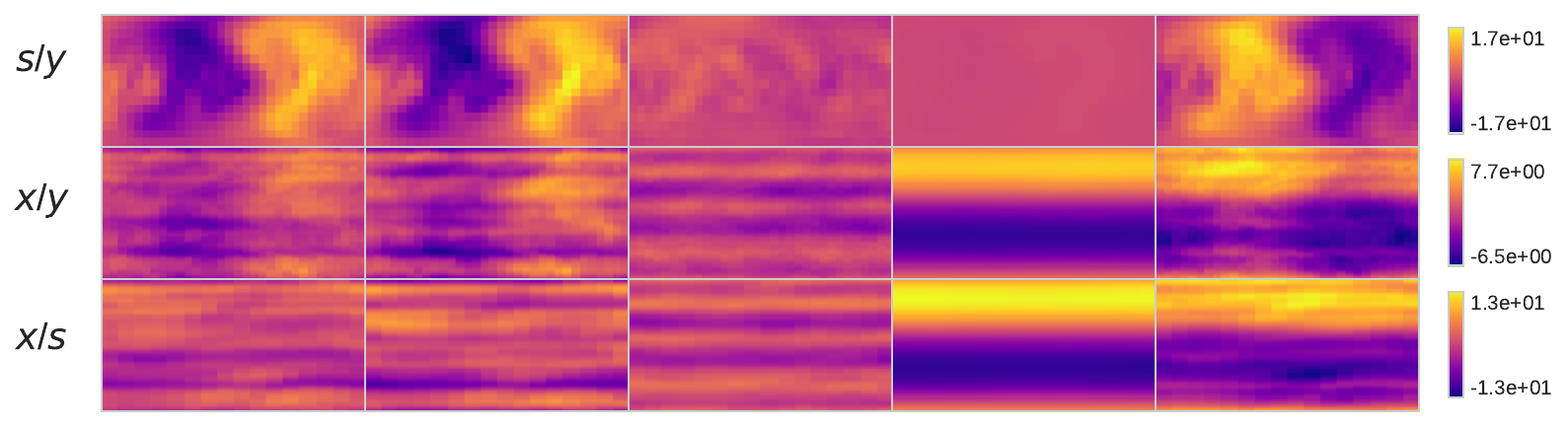} \\
    \end{tabular}
    \caption{
    Extra reconstructions for the 3D electrostatic potential $\bm{\phi}$. $\text{CR}=\sim1{,}000\times$.
    Each row is a different trajectory at timestep $176.4R/V_\mathrm{r}$. Columns match \cref{fig:extra_df}.
    Aspect ratio is set to 2 and does not represent the physical one.
    }
    \label{fig:extra_phi}
\end{figure}

\end{document}